\documentstyle[prd,aps,eqsecnum]{revtex}
\input psfig 
\begin{document}

\newcommand{\WH}{\mbox{{\tiny $W\!H$}}}
\newcommand{\NB}{\mbox{{\tiny $N\!B$}}}
\newcommand{\Vil}{\mbox{{\tiny $Vil$}}}
\newcommand{\EH}{\mbox{{\tiny $E\!H$}}}
\newcommand{\C}{\mbox{{\rm C}}\!\!\!
 \mbox{\raisebox{.4 ex}{\scriptsize  $|$}}\ }
\newcommand{\R}{\mbox{{\rm R}\hspace{.05 ex}}\!\!\!\!\! 
 \mbox{\raisebox{.4 ex}{{\scriptsize $|$}}}\ \,}
\newcommand{\dd}{\mbox{d}}
\newcommand{\bpm}{\beta_{\pm}}
\newcommand{\LL}{{\cal L}}
\newcommand{\HH}{{\cal H}}
\newcommand{\Ak}{{\cal A}}
\newcommand{\CC}{{\cal C}}
\newcommand{\RR}{{\cal R}}
\newcommand{\KK}{{\cal K}}
\newcommand{\II}{{\cal J}}

\newcommand{\slambda}{\ \, \sim^{^{\mbox{\hspace{-.45 cm}{\scriptsize $\lambda\! \to \!0$}}}}\,}

\newcommand{\tlambda}{\ \, \to^{^{\mbox{\hspace{-.45 cm}{\scriptsize $\lambda\! \to \!0$}}}}\,}

\newcommand{\skappa}{\ \ \, \sim^{^{\mbox{\hspace{-.5 cm}{\scriptsize $\kappa\! \to \! \infty$}}}}\,}

\newcommand{\skappabeta}{\ \,\sim^{^{\mbox{\hspace{-.5 cm}{\scriptsize $\kappa\! \to \!\infty$}}}}_{_{\mbox{\mbox{\hspace{-.45 cm}{\scriptsize $\beta \! \to \!0$}}}}}}

\newcommand{\stau}{\ \, \sim^{^{\mbox{\hspace{-.45 cm}{\scriptsize $\tau\! \to \!0$}}}}\ \,}

\newcommand{\ptau}{\ \, \propto^{^{\mbox{\hspace{-.4 cm}{\scriptsize $\tau\! \to \!0$}}}}\,}

\newcommand{\shbar}{\ \, \sim^{^{\mbox{\hspace{-.45 cm}{\scriptsize $\hbar\! \to \!0$}}}}\,}

\newcommand{\sm}{\ \  \sim^{^{\mbox{\hspace{-.5 cm}{\scriptsize $m\! \to \!0$}}}}\,}

\title{Physical states of Bianchi type IX quantum cosmologies described by the Chern-Simons functional}
\author{Robert Graham and Robert Paternoga}
\address{Fachbereich Physik, Universit\"at-Gesamthochschule Essen,
 45117 Essen, Germany}
\maketitle

\begin{abstract}
A class of exact solutions of the Wheeler-DeWitt equation for diagonal Bianchi type IX cosmologies with cosmological constant is derived in the metric representation. This class consists of all the ``topological solutions'' which are associated with the Bianchi type IX reduction of the Chern-Simons functional in Ashtekar variables. The different solutions within the class arise from the topologically inequivalent choices of the integration contours in the transformation from the Ashtekar representation to the metric representation. We show how the saddle-points of the reduced Chern-Simons functional generate a complete basis of such integration contours and the associated solutions. Among the solutions we identify two, which, semi-classically, satisfy the boundary conditions proposed by Vilenkin and by Hartle and Hawking, respectively. In the limit of vanishing cosmological constant our solutions reduce to a class found earlier in special fermion sectors of supersymmetric Bianchi type IX models.
\end{abstract}

\section{Introduction}

Three decades after the first exploratory steps were made \cite{1,2} canonical quantum gravity continues as a vigorous program of fundamental research. New life has been breathed into this field about a decade ago by Ashtekar's discovery \cite{4} of a representation of general relativity in terms of new variables, which render the Hamiltonian and diffeomorphism constraints more tractable. The developments in nonperturbative canonical quantum gravity which followed this advance \cite{5,6,7,8} justify a reasonable hope that a mathematically consistent quantum gravity might be attainable. Still, the program is far from completed and many facets of the theory remain to be explored. One such facet is the relation between the metric representation and Ashtekar's representation. This question is far from trivial: Ashtekar's formulation starts out with complexified general relativity and suitable reality conditions on the new variables have to be imposed at the end. The metric representation, on the other hand, stays within the domain of the real theory all along. Thus it is not clear, a priori, whether there is a unique one to one relation between the complexified theory and the real theory. Can, e.g., a single solution of the Wheeler-DeWitt equation in Ashtekar's variables before imposing reality conditions give rise to several mathematically and physically distinct solutions in metric variables?

In the present paper we wish to examine this question in the framework of the minisuperspace model of diagonal Bianchi type IX with a non-vanishing cosmological constant.

Kodama \cite{9} and Brencowe \cite{10} have found a simple solution of the basic constraints of quantum gravity with cosmological term in Ashtekar's variables in the form of an exponential of the Chern-Simons functional. By projecting to Bianchi type IX geometries one also obtains a solution for the minisuperspace model \cite{8}.

In the present paper we shall start from the Wheeler-DeWitt equation for the minisuperspace model in the metric representation and a specific choice of operator ordering (which is gleaned from the special operator ordering appearing naturally if supergravity is used as a starting point \cite{11,12,12a,13}). The Ashtekar representation is then introduced only on the quantum level as a mathematical device, like a Laplace-transform, to simplify the equations. In fact, the representation is introduced as a kind of complexified momentum-representation, in which Ashtekar's variables are the complexified canonically conjugate momenta of the inverse triad corresponding to the Bianchi type IX 3-metric. The integration contour in the complex manifold spanned by these momentum variables may be chosen quite freely within the requirements of convergence and the vanishing of boundary terms in partial integrations. Integration contours which can be deformed into each other while satisfying these requirements are topologically equivalent. However, a given solution in Ashtekar variables may admit topologically inequivalent choices of integration contours. Such a solution in Ashtekar variables may then correspond to several mathematically and physically distinct solutions in the metric representation. In fact, we shall show that this happens for the Chern-Simons topological solution in the diagonal Bianchi type IX minisuperspace model. We find that five topologically inequivalent integration contours over the Ashtekar variables exist, which are organized by five distinct saddle-points of the reduced Chern-Simons functional and the accompanying paths of steepest ascent and descent. These findings raise the interesting question, whether similar results may also be obtained in the full theory. The answer to this question is not obvious, because the enlargement of the configuration space, in principle, could render integration contours topologically equivalent, which appear as topologically inequivalent when projected on the minisuperspace under investigation.

While this general question transcends our minisuperspace framework and must be left open here, our results for the minisuperspace model yield several new exact solutions in metric variables with non-vanishing cosmological constant. These solutions turn out to be of interest in their own right. To simplify their discussion we restrict ourselves to the physically more interesting case $\Lambda >0$ throughout this paper and postpone the examination of the case $\Lambda <0$ to a future work. We discuss the asymptotic limits of the solutions for $\hbar \to 0$ and $\Lambda \to 0$ and show that, at least semi-classically, two of them satisfy the cosmological boundary conditions proposed by Vilenkin \cite{14} and by Hartle and Hawking \cite{15,16}, respectively. Furthermore, we show that it is just the no-boundary state which additionally fulfills a physically well-motivated normalizability condition.

The remainder of this paper is organized as follows: In section \ref{1} we establish our notation, put down the Wheeler-DeWitt equation in the adopted operator ordering, and briefly list the five exact solutions for vanishing cosmological constant which are known from recent work on Bianchi type IX supergravity. In section \ref{2} the Ashtekar representation is introduced as a complexified momentum-representation, and the resulting Chern-Simons solution of the Wheeler-DeWitt operator is given. In section \ref{3} that solution is transformed to the metric representation along five topologically distinct integration contours, which establishes a basis of five linearly independent solutions which are all generated by the Chern-Simons solution. Their asymptotic behavior for $\hbar \to 0$, $\Lambda \to 0$, or $\Lambda \, a^{2} \to \infty$ (where $a$ is the scale parameter and $\Lambda$ the cosmological constant) is studied in section \ref{4}. Here, in addition, we establish the relation of two of these solutions to those picked out by the boundary or no-boundary conditions of Vilenkin, and Hartle and Hawking, respectively. In the last section we draw some conclusions and indicate how our results may also be used to establish certain limiting forms of Bianchi type IX models coupled to a scalar matter field with very small mass.          

\section{Metric representation}\label{1}

\subsection{Wheeler-DeWitt equation}

The purpose of this section is to establish some notation and to derive the Wheeler-DeWitt equation for the Bianchi type IX model in the metric representation in a specific factor-ordering. We start from the Einstein-Hilbert action with a cosmological constant $\Lambda$ 

\begin{equation}\label{1.1}
{\cal S}_{\EH}\ [g_{\mu \nu}]\ =\ \frac{1}{16 \pi}\ \int\limits_{{\cal M}} \dd ^{4} x \sqrt{-g}\ \Bigl( \RR-2 \Lambda \Bigr )
\end{equation}
\\
where a possible boundary term has been omitted, since such a term will not contribute to the resulting Lagrangian density. In (\ref{1.1}) the action integral is taken over the 4-dimensional space-time manifold of the Universe, while $g$ and $\RR$ are the determinant and the Ricci-scalar of the 4-metric {\boldmath $g$}$\,=(g_{\mu \nu})$, respectively. Performing the ADM space-time split a {\em lapse}-function $N$, a {\em shift}-vector $N^{i}$ and the 3-metric $h_{i j}$ of the spatial slice are introduced in the usual way \cite{17}. Then  the action (\ref{1.1}) takes the form

\begin{equation}\label{1.2}
{\cal S}_{\EH} =\int \dd t \int \dd ^{3}x\, \frac{N \sqrt{h}}{16\, \pi} \  \Bigl (\, ^{3}\RR-K^{2}+K_{i j} K^{i j}-2 \Lambda \Bigr)\ ,
\end{equation}
\\
where $h=\det\, (h_{i j} )$, $^{3}\RR$ is the curvature scalar of the spatial manifold, and $K_{i j}$ is the extrinsic curvature tensor. The spatial homogeneity manifests itself in the existence of infinitesimal coordinate transformations $x'^{i} = x^{i}+\varepsilon^{j}\,  \xi^{i}_{(j)}(x)$, which leave tensors on the 3-manifold form-invariant \cite{18}. In the Bianchi type IX case the algebra of the Killing-vectors with $\vec \xi_{(i)}(x)=\xi^{j}_{(i)} \frac{\partial}{\partial x^{j}}$ is chosen to be 

\begin{displaymath}
\lbrack \vec \xi_{(i)}, \vec \xi_{(j)} \rbrack =  \varepsilon_{i j k}  \ \vec \xi_{(k)}\ ,
\end{displaymath}
\\
implying $\dd\,\tilde \omega^{i} = \frac{1}{2}\  \varepsilon_{i j k}\ \tilde \omega^{j} \wedge\tilde\omega^{k}$ for the invariant basis $\tilde \omega^{i}$ \cite{18}. Using this special basis in the following, the components of all tensors on the 3-manifold become functions of time $t$ only.
Let us now consider the Bianchi type IX form \cite{17,18,19,20,21,22a}

\begin{equation} \label{1.4}
h_{i j}(t)=  e^{2 \alpha(t)}\ (e^{2 \mbox{{\footnotesize $\beta$}} (t)})_{i j}
\end{equation}
\\
for the 3-metric, but with restriction to the diagonal case. Then  $(\beta_{i j})$ is a diagonal, traceless $3 \times 3$-matrix, which can be parameterized as 

\begin{equation}\label{1.5}
(\beta_{i j})=\mbox{diag}(\beta_{+}+\sqrt{3}\  \beta_{-}, \beta_{+}-\sqrt{3}\  \beta_{-}, -2\, \beta_{+})\ .
\end{equation}
\\
The action (\ref{1.2}) yields a Lagrangian

\begin{equation}\label{1.6}
\LL = 2\, \pi N e^{3 \alpha}\ \Bigl (\, ^{3}\RR-K^{2}+K_{i j} K^{i j}-2 \Lambda \Bigr )\ , 
\end{equation}
\\
where the spatial integration has been carried out using $\int \tilde \omega^{1} \wedge\, \tilde \omega^{2} \wedge\, \tilde \omega^{3}= 2 \,(4 \pi)^{2}$, and $^{3}\RR$ and $K_{i j}$ have to be expressed in terms of the new metric variables $\{\alpha, \bpm \}$ and their time-derivatives. It turns out, that $\LL$ is independent of $\dot N$ and $\dot N^{i}$, so the conjugate momenta satisfy the primary constraints

\begin{equation}\label{1.7}
\pi^{N}\!:=\frac{\partial \LL}{\partial \dot N} \equiv 0\qquad ,\qquad \pi^{i}\!:=\frac{\partial \LL}{\partial \dot N_{i}} \equiv 0\ ,
\end{equation}
\\
i.e. they vanish identically. The preservation of the primary constraints in time leads to the secondary constraints

\begin{equation}
\frac{\partial \LL}{\partial N} \equiv 0 \equiv
\frac{\partial \LL}{\partial N^{i}}\ ,
\end{equation}
\\
the first of which is the Hamiltonian constraint, while the second one, in the present case, is solved by taking $N^{i} \equiv 0$. Therefore  the 4-metric is now of the form

\begin{equation}\label{1.8}
\mbox{{\boldmath $g$}}=-N^{2}(t)\,\dd t^{2}+h_{i j}(t)\   \tilde \omega^{i} \otimes \tilde \omega^{j} \ .
\end{equation}
\\
A straightforward Legendre transform yields the Hamiltonian

\begin{equation}\label{1.9}
\HH=\frac{N \,  e^{-3 \alpha}}{48 \, \pi}  H\ ,\ \mbox{where}
\ \ H= -\pi^{\alpha^{2}}+\pi^{+^{2}}+\pi^{-^{2}} +e^{4 \alpha}\, U(\bpm) + 3\,(8 \pi)^{2}\, \Lambda \,e^{6 \alpha} \ .
\end{equation}
Here the momenta
\begin{equation}\label{1.10}
\pi^{\alpha}=\frac{\partial \LL}{\partial \dot \alpha}\ \ \ ,\ \ \ 
\pi^{\pm}=\frac{\partial \LL}{\partial \dot \bpm} 
\end{equation}
\\
have been introduced, which are connected to the generalized velocities as follows:

\begin{equation}\label{1.11}
\dot \alpha =- \frac{N}{24 \pi} e^{-3 \alpha}\  \pi^{\alpha}\ \ \ ,\ \ \  \dot \bpm = \frac{N}{24 \pi} e^{-3 \alpha}\  \pi^{\pm}\ .
\end{equation}
\\
Furthermore, the curvature-potential $U(\bpm)$ has been  defined as

\begin{equation}\label{1.12}
U(\bpm):=-6\,(4 \pi)^{2}\  e^{2 \alpha}\ ^{3} \RR =3\, (4 \pi)^{2} \  \mbox{Tr} \left (e^{4 \mbox{{\footnotesize $\beta$}}}-2\,e^{-2 \mbox{{\footnotesize $\beta$}}} \right )\ .
\end{equation}
\\
The Hamiltonian constraint now simply reads $H \equiv 0$.

To quantize the model, one may seek for solutions of the Schr\"odinger equation $\mbox{$\HH$}\  |\Psi\rangle = E\  |\Psi\rangle $, 
where $\HH$ is now interpreted as a self-adjoint operator over a suitably defined Hilbert space of wavefunctions $|\Psi\rangle$. The Hamiltonian constraint, which must be satisfied by the physical states,  then implies a restriction to zero-energy states, yielding the Wheeler-DeWitt equation

\begin{equation}\label{1.13}
H\  |\Psi\rangle = 0\ . 
\end{equation}
\\
The quantized version of the Hamiltonian (\ref{1.9}), which is obtained from the usual rules of canonical quantization, suffers from the well-known ambiguity in the choice of the factor-ordering. We shall resolve this ambiguity in such a way that a class of simple semi-classical solutions of (\ref{1.13}) (to be given in the following section), whose existence and form, due to their semi-classical nature, are independent of the factor-ordering, become {\em exact} solutions. This will be achieved through the following non-standard procedure: It is easily checked that the classical Hamiltonian (\ref{1.9}) can be written in the form\footnote{These definitions, and the resulting factor-ordering, are suggested by the existence of a supersymmetric extension of this form of the Hamiltonian \cite{11,12,13}. The factor-ordering chosen here is not contained in the class considered in \cite{22}.} 

\begin{equation}\label{1.14}
H = \Bigl \lbrack \, i \pi^{\alpha}-\Phi_{,\alpha}\,  \Bigr \rbrack \Bigl \lbrack \, i \pi^{\alpha}+\Phi_{,\alpha}\,  \Bigr \rbrack-\Bigl \lbrack \, i \pi^{+}-\Phi_{,+} \Bigr\rbrack \Bigl \lbrack \, i \pi^{+}
+\Phi_{,+} \Bigr \rbrack - \Bigl \lbrack \, i \pi^{-}-\Phi_{,-} \Bigr \rbrack \Bigl \lbrack \, i \pi^{-}+\Phi_{,-} \Bigr \rbrack+3\, (8 \pi)^{2}\,\Lambda\,\,e^{6 \alpha} \ ,
\end{equation}
\\
where $\Phi$ is defined as 

\begin{equation}
\Phi:=2 \pi\,e^{2 \alpha}\, \mbox{Tr}\, e^{2 \mbox{{\footnotesize $\beta$}} }\ .
\end{equation}
\\
Assuming now canonical commutation relations
$\lbrack \alpha, \pi^{\alpha} \rbrack=\lbrack \bpm,\pi^{\pm} \rbrack=i \hbar$, 
the momenta in the $\{\alpha, \bpm \}$-representation may be expressed as
$ \pi^{\alpha}=-i\hbar\ \partial _{\alpha},\  
\pi^{\pm}=-i\hbar\ \partial _{\pm}\ $,
and $H$ becomes

\begin{equation}\label{1.15}
H = \Bigl \lbrack \, \hbar \partial_{\alpha}-\Phi_{,\alpha}\,  \Bigr \rbrack \Bigl \lbrack \, \hbar \partial_{\alpha}+\Phi_{,\alpha}\,  \Bigr \rbrack-\Bigl \lbrack \, \hbar \partial_{+}-\Phi_{,+} \Bigr\rbrack \Bigl \lbrack \, \hbar \partial_{+}
+\Phi_{,+} \Bigr \rbrack - \Bigl \lbrack \, \hbar \partial_{-}-\Phi_{,-} \Bigr \rbrack \Bigl \lbrack \, \hbar \partial_{-}+\Phi_{,-} \Bigr \rbrack+3\,(8 \pi)^{2}\,\Lambda  \,e^{6 \alpha} \ .
\end{equation}
\\
Finally, this corresponds to a Wheeler-DeWitt equation

\begin{equation}\label{1.16}
\biggl\{
\hbar^{2} \left \lbrack  \partial^{2}_{\alpha}-\partial^{2}_{+}-\partial^{2}_{-}\, \right \rbrack -12 \,\hbar \Phi +e^{4 \alpha}\   U(\bpm\!)+
3\, (8 \pi)^{2}\, \Lambda\,e^{6 \alpha} 
\biggr \}\,\Psi(\alpha,\bpm;\Lambda)=0
\end{equation}
\\
in the metric representation. The non-standard term $-12\, \hbar \Phi$ is a quantum correction to the classical ``potential'' $e^{4 \alpha}\, U(\bpm)$ and appears as a result of our choice of operator ordering in equation (\ref{1.15}).

\subsection{Solutions without cosmological constant}

For completeness, and because they will play some role in the discussion of the general solutions in the case $\Lambda \not=0$, we shall now present a derivation of the five known solutions of the Wheeler-DeWitt equation (\ref{1.16}) without cosmological constant \cite{11,12,12a,13}. 

An obvious, exact solution of the Wheeler-DeWitt equation with vanishing $\Lambda$ can be extracted from the form (\ref{1.15}) immediately \cite{11}:

\begin{equation}\label{1.17}
\Psi_{\WH}^{0}:=\exp \left \lbrack -\frac{\Phi}{\hbar} \right \rbrack \ .
\end{equation}
\\
In the classical limit $\hbar \to 0$ one can interpret $S=i \Phi$ as the Euclidean action of this wavefunction. As known from Hamilton-Jacobi theory the derivatives of the action with respect to the generalized coordinates play the role of the generalized momenta, so the classical trajectories may be computed via (\ref{1.11}) from

\begin{equation}\label{1.18}
\frac {\dd \alpha}{i\,N \dd t}=-\frac{e^{-3 \alpha}}{24 \pi}\   \Phi_{, \alpha}\ \ , \ \ 
\frac {\dd \bpm}{i\,N \dd t}=\frac{e^{-3 \alpha}}{24 \pi}\   \Phi_{, \pm}\ .
\end{equation}
\\
Clearly, because $\Phi\, \epsilon\, \R$, no real solutions to (\ref{1.18}) exist, but one can introduce a new parameter of imaginary time $\dd \tau:=i N \dd t$ and look for Euclidean solutions, corresponding to positive definite 4-manifolds via (\ref{1.8}). It can be shown that the solutions of (\ref{1.18}) in this Euclidean regime form a two-parameter family of classical Universes, which become all asymptotically flat and isotropic in the limit $a \to \infty$, where $a:=2\, e^{\alpha}$ is the average scale factor.
In fact, (\ref{1.17}) is the well-known ``wormhole-state'' of the Bianchi type IX model. 

To derive other solutions of the model without cosmological constant we will first subject the Hamiltonian (\ref{1.15}) to the similarity-transformation

\begin{equation}\label{1.19}
H=e^{ -\frac{\Phi}{\hbar}}\  H'\ e^{ \frac{\Phi}{\hbar}} \ ,\ \ |\Psi\rangle=\ e^{ -\frac{\Phi}{\hbar}} \ |\Psi'\rangle
\end{equation}
\\
which yields

\begin{equation}\label{1.20}
H'=\hbar^2\,  \Bigl \lbrack \partial_{\alpha}^{2}- \partial_{+}^{2}-\partial_{-}^{2} \Bigr \rbrack-2\, \hbar \Bigl \lbrack \Phi_{,\alpha}\,\partial_{\alpha} -\Phi_{,+}\,  \partial_{+}-\Phi_{,-}\,\partial_{-} \Bigr \rbrack+
3\,\pi^{2} a^{6} \Lambda\ .
\end{equation}
\\
Moreover, new variables playing the role of the inverse triad are introduced 

\begin{equation}\label{1.21}
\sigma_{i}:=\frac{\pi}{\hbar}\, a^{2}\,e^{-\beta_{i}}\, (\geq 0)\ ,\ \   i\,\epsilon\,\{1,2,3\}\ , 
\end{equation}
\\
where $\beta_{i}$ are the diagonal elements of the matrix $(\beta_{i j})$. Denoting derivatives with respect to these new variables with
$\partial_{i}:= \frac{\partial}{\partial \sigma_{i}}$ and rescaling the cosmological constant into $\lambda=\frac{\hbar \Lambda}{6 \pi}$, 
one finds for the Hamiltonian in the $\sigma_{i}$-representation:

\begin{equation}\label{1.23}
 H'=12\, \hbar^{2}\ \sum_{i=1}^{3}\  H'_{i} \ ,\ \ 
\mbox{where}\ \ 
 H'_{i}:=\sigma_{j} \sigma_{k} \left \lbrack\, \partial_{j} \partial_{k}-\partial_{i} +\frac{\lambda}{2}\,  \sigma_{i} \,\right \rbrack\ \ ,\ \varepsilon_{i j k}=1\ .
\end{equation}
\\
Two important features of $H'$ in this representation should be mentioned:

\begin{itemize}

\item $H'$ is a sum of three terms, producing each other by cyclic permutation of the indices.

\item $H'$ is invariant under converting two of the $\sigma_{i}$ into their negatives.

\end{itemize}
The first of these properties suggests to look for solutions by solving the reduced equation 

\begin{equation}\label{1.24}
\sigma_{1} \sigma_{2} \,\Bigl \lbrack \partial_{1} \partial_{2}-\partial_{3} \Bigr \rbrack\,\Psi'=0
\end{equation}
\\
with a $\sigma_{i}$-symmetric function $\Psi'$. The simplest solution to this ansatz turns out to be

\begin{equation}\label{1.25}
\Psi'^{0}_{\NB}:=\exp \Bigl \lbrack  
\sigma_{1}+\sigma_{2}+\sigma_{3}  \Bigr \rbrack\ ,
\end{equation}
\\
which by multiplication with the wormhole-state gives a second solution, the ``Hartle-Hawking state'' \cite{12,12a,13}. This name can be justified by discussing the equations for the classical trajectories analogous to (\ref{1.18}). There it turns out that each member of the two-parameter family of solutions describes a regular (and isotropic) Universe in the limit $a \to 0$, i.e. {\em in this sense} $\Psi^{0}_{\NB}$ satisfies the ``no-boundary'' proposal.

Three further solutions to the Wheeler-DeWitt equation occur just because of the second property of the $\sigma_{i}$-representation of $H'$, namely \cite{13}

\begin{equation}\label{1.26}
\Psi'^{0}_{i}:=\exp \Bigl \lbrack  
\sigma_{i}-\sigma_{j}-\sigma_{k} \Bigr \rbrack\ ,\ \ 
\varepsilon_{i j k}=1\ .
\end{equation}
\\
These ``asymmetric solutions'' create classical Universes which turn out to be asymptotically three dimensional: In the limit $a \to \infty$ the $i$-th dimension is curved to zero, while the remaining manifold becomes an asymtotically flat, spatially two dimensional wormhole.

We shall add a few remarks on these states concerning their {\em normalizability}. An investigation of the five wavefunctions reveals that $\Psi_{\WH}^{0}$ and the three asymmetric solutions $\Psi_{i}^{0}$ are {\em bounded} functions on minisuperspace, whereas $\Psi_{\NB}^{0}$ is unbounded (in fact, it grows super-exponentially for $\alpha \to +\infty$, when $\bpm$ is kept fixed and small). In the limit $\alpha \to -\infty$, i.e. considering  vanishing scale factors, all five solutions approach unity, and thus a normalization integral over the full $\{ \alpha, \bpm \}$-space diverges in any case. However, the four bounded solutions $\Psi_{\WH}^{0}$ and $\Psi_{i}^{0}$ may at least be called normalizable in the distributional sense. Introducing a suitable integration weight they will even become normalizable in the usual sense.

\section{Ashtekar representation}\label{2}

\subsection{Wheeler-DeWitt equation in Ashtekar's variables}

The transformation of the Wheeler-DeWitt equation (\ref{1.16}) into the Ashtekar representation \cite{23,24} has been well-prepared during the last section by introducing the new triad variables $\sigma_{i}$ via (\ref{1.21}). All that remains to be done now is to transform (\ref{1.23}) to the generalized momenta $\Ak_{i}$ conjugate to $\sigma_{i}$ by performing a suitable  Fourier-transformation.

It will be crucial for all that follows that we shall choose {\em not} a standard Fourier-transformation which is carried out along the real axes, but a generalized, complexified Fourier-transformation defined as\footnote{Since we are dealing with complex-valued quantities $\Ak_{i}$ anyway, the standard ``$i$'' has been absorbed in these new variables.}

\begin{equation}\label{2.1}
\Psi'(\vec\sigma)=\int\limits_{\Sigma} \mbox{d}^{3} {\cal A}\ e^{\vec \sigma \cdot \vec {\cal A}}\ \tilde \Psi'(\vec {\cal A})\ .
\end{equation}
\\
Here we assume that $\Sigma \subset \C^{3}$ is a smooth, three dimensional manifold with the following properties:

\vspace{.3 cm}

($i$)$\,$ The integrand of (\ref{2.1}) and its first and second derivatives with respect to $\Ak_{i}$ vanish at the border $\partial \Sigma$ of $\Sigma$.

\vspace{.2 cm}

($ii$) $\Sigma$ is completely contained within a domain where $\tilde \Psi '(\vec \Ak)$ is a holomorphic function of $\Ak_{i}$.

\vspace{.3 cm}

\noindent
Assuming these properties of $\Sigma$ it is possible to convert  $\sigma_{i}$ and $\partial_{i}$ to the new variables: It turns out that they obey exactly the classical conversion rules

\begin{equation}\label{2.2}
\partial_{i} \to \Ak_{i}\ \ ,\ \ \sigma_{i} \to - \tilde \partial_{i}\ ,
\end{equation}
\\
where for brevity $\tilde \partial_{i}:=\frac{\partial}{\partial \Ak_{i}}$, i.e. the form of the Fourier-transformed equation is independent of the choice of $\Sigma$ and unaffected by our complexification procedure!

The transformed equation (\ref{1.23}) now reads

\begin{equation}\label{2.3}
\tilde H'\  \tilde \Psi'=0\ ,\ \ \mbox{where}\ \ 
\tilde H'=\sum_{i=1}^{3}\ \tilde H'_{i}\ \ ,
\ \ \tilde H'_{i}:=
\tilde \partial_{j}\tilde \partial_{k} \left \lbrack\,
{\cal A}_{j} {\cal A}_{k}-{\cal A}_{i}-\frac{\lambda}{2}\,\tilde \partial_{i} 
\,\right \rbrack \ ,
\end{equation} 
\\
which we will refer to as the Wheeler-DeWitt equation in the Ashtekar representation.

\subsection{Solution with cosmological constant}

An exact solution of the Wheeler-DeWitt equation with non-vanishing cosmological constant can now be easily constructed: As in section II.B we try to make vanish each term of the Hamiltonian of (\ref{2.3}) seperately, i.e. we seek for a solution obeying the three equations 

\begin{equation}\label{2.4}
\tilde H'_{i}\ \tilde \Psi'=
\tilde \partial_{j}\tilde \partial_{k} \left \lbrack\,
{\cal A}_{j} {\cal A}_{k}-{\cal A}_{i}-\frac{\lambda}{2}\,\tilde \partial_{i} 
\,\right \rbrack \ \tilde \Psi'=0\ .
\end{equation}
\\
Furthermore, one may even try to solve

\begin{equation}\label{2.5}
\left \lbrack
{\cal A}_{j} {\cal A}_{k}-{\cal A}_{i}-\frac{\lambda}{2}\,\tilde \partial_{i} 
\right \rbrack \ \tilde \Psi'=0
\end{equation}
\\
for each set of $\{i, j, k\}=\{1, 2, 3\}$ simultaneously. This requirement gives the unique solution

\begin{equation}\label{2.6}
\tilde \Psi'(\vec \Ak)=\exp \left \lbrack
\frac{1}{\lambda}\,\left (\, 2\,{\cal A}_{1} {\cal A}_{2} {\cal A}_{3}- \vec \Ak^{\,2} \,\right ) \right \rbrack \ .
\end{equation}
\\
This solution is even known for the general spatially inhomogeneous case, where the exponent is given by the Chern-Simons functional \cite{9,10}. In \cite{24} the semi-classical content of the wave function (\ref{2.6}) was analyzed within the Ashtekar representation. Here we shall be interested, instead, in its transformation to the metric representation. In \cite{9} the transformation back to metric variables was also attempted, but without success. In the following, we shall make use of the freedom in the choice of contours in the generalized Fourier-transformation (\ref{2.1}) to derive {\em several different} solutions in the metric representation which are all generated from (\ref{2.6}) by the choice of topologically inequivalent contours.

\section{Transformation to the metric representation}\label{3}

\subsection{General form of the transformation}

By defintion, the generalized Fourier-transform (\ref{2.1}) of (\ref{2.6}) represents an exact solution of the Wheeler-DeWitt operator (\ref{1.23}) in the $\sigma_{i}$-representation: Each wavefunction of the form

\begin{equation}\label{3.1}
\Psi'(\vec \sigma)= \int\limits_{\Sigma} \dd^{3} \Ak \ \exp \left \lbrack \ \frac{2}{\lambda} \left(\vec \Ak \cdot \vec \kappa -\frac{1}{2} \vec \Ak^{\ 2} +\Ak_{1}\Ak_{2}\Ak_{3} \right )\  \right \rbrack \ ,
\end{equation}
\\
where $\Sigma$ is chosen to imply a sufficiently large fall-off for the integrand on $\partial \Sigma$, is a solution for the Bianchi type IX model with cosmological constant. In (\ref{3.1}) new variables

\begin{equation}\label{3.2}
\kappa_{i}:=\frac{1}{2}\,\lambda\,\sigma_{i}=\frac{1}{12}\,\Lambda\, a^{2} e^{-\beta_{i}}
\end{equation}
\\
have been introduced. We will show that there exist {\em several} manifolds $\Sigma$ satisfying the conditions ($i$) and ($ii$), corresponding to {\em different} solutions in the metric representation, so obviously, due to the existence of several topologically inequivalent contours, the generalized Fourier-transformation is not unique. It will turn out that for the space of exact solutions defined via (\ref{3.1}) the number and location of the saddle-points of the integrand's exponent

\begin{equation}\label{3.3}
F(\vec \Ak,\vec \kappa):=\vec \Ak \cdot \vec \kappa -\frac{1}{2} \vec \Ak^{\ 2} +\Ak_{1}\Ak_{2}\Ak_{3}
\end{equation}
\\
will play an essential role. While for the asymptotic form of the solutions  obtained from (\ref{3.1}) by the saddle-point method the importance of the saddle-points is obvious and well-known \cite{25,26}, their importance for the {\em exact} solutions is a surprise, which arises because of the freedom in the choice of the integration contours. These saddle-points are determined by the equations

\begin{equation}\label{3.4}
\frac{\partial F}{\partial \Ak_{i}}=0\ \  \Leftrightarrow\ \ \kappa_{i}-\Ak_{i}+\Ak_{j}\Ak_{k}=0\ \ ,\ \ \varepsilon_{i j k}=1
\end{equation}
\\
which in the case $\kappa_{1} \not= \kappa_{2} \Rightarrow \Ak_{3} \not= \pm 1$ may be rewritten in the form

\begin{eqnarray}
\Ak_{1}&=&\kappa_{1}+\Ak_{2}\Ak_{3} \ ,\label{3.5}\\[0.4 cm]
\Ak_{2}\,(1-\Ak_{3}^{2})&=&\kappa_{2}+\kappa_{1}\, \Ak_{3}\ ,\label{3.6}\\[0.4 cm]
(\Ak_{3}-\kappa_{3})\,(1- \Ak_{3}^{2})^{2}\!\!&=&(\kappa_{1}+\kappa_{2}\, \Ak_{3})(\kappa_{2}+\kappa_{1}\, \Ak_{3})\ .\label{3.7}
\end{eqnarray}
\\
The third equation (\ref{3.7}) is of fifth order and so yields five different solutions, which correspond to five different saddle-points via (\ref{3.6}) and (\ref{3.5}).

One may show that also in the remaining case $\kappa_{1}=\kappa_{2}$ (\ref{3.4}) has five different solutions with $\Ak_{3}$-components still obeying (\ref{3.7}).
The fact that there always exist five saddle-points of the integrand's exponent will be shown to result in a five dimensional space of solutions (\ref{3.1}).

For concreteness, let us assume a special representation of sucessive integrations 

\begin{eqnarray}\label{3.8}
\Psi'(\vec\sigma)
&=&
\int\limits_{\Gamma_{3}} \dd\Ak_{3}\  \!\!
\int\limits_{\Gamma_{2}(\Ak_{3})} \!\!\!\dd\Ak_{2}\  \!\!\!\!
\int\limits_{\Gamma_{1}(\Ak_{2},\Ak_{3})} \!\!\!\!\!\!\dd\Ak_{1}\  
\ \exp \left \lbrack \,\frac{2}{\lambda}\, F(\vec \Ak,\vec \kappa) \,\right \rbrack\nonumber\\[0.6 cm]
&=&
\int\limits_{\Gamma_{3}} \dd\Ak_{3}\  \!\!
\int\limits_{\Gamma_{2}(\Ak_{3})} \!\!\!\dd\Ak_{2}\ 
\exp\left\lbrack\,\frac{1}{\lambda}\,\biggl(2 \Ak_{2}\kappa_{2}+2 \Ak_{3}\kappa_{3}-\Ak_{2}^{2}-\Ak_{3}^{2}\biggr)\,\right\rbrack\nonumber\\[0.5 cm]
& &\ \ \ \ \ \ 
\mbox{\raisebox{0.3 ex}{$\cdot$ \hspace{-.7cm}}}
\int\limits_{\Gamma_{1}(\Ak_{2},\Ak_{3})} \!\!\!\!\!\dd\Ak_{1}\ \,
\exp\left\lbrack\,\frac{1}{\lambda}\biggl(-\Ak_{1}^{2}+2 \Ak_{1}\,( \kappa_{1}+\Ak_{2}\Ak_{3})\biggr)\,\right\rbrack\ ,
\end{eqnarray}
\\
which one may show to be of no restriction to the general case. The one dimensional Gaussian $\Ak_{1}$-integral just has a single saddle-point located at 

\begin{equation}\label{3.9}
\Ak_{1}=\kappa_{1}+\Ak_{2}\Ak_{3}\ .
\end{equation}
\\
Up to a factor that may depend on $\lambda$, but that will be absorbed in a proportionality sign ``$\propto$'' in the following, there is only one non-trivial value this integral can take: Since the integrand has to vanish at the ends of $\Gamma_{1}$ each integration curve can be deformed into the curve of steepest descent, which in the new coordinate

\begin{equation}\label{3.10}
\Ak_{1}'=\Ak_{1}-(\kappa_{1}+\Ak_{2}\Ak_{3})
\end{equation}
\\ 
turns out to be simply the real axis. So the $\Ak_{1}$-integration yields

\begin{equation}\label{3.11}
\int\limits_{-\infty}^{+\infty}\,
\dd\Ak_{1}'\ \exp \left \lbrack\,\frac{1}{\lambda}\,\left(-\Ak_{1}'^{2} +\bigl (\kappa_{1}+\Ak_{2}\Ak_{3} \bigr)^{2}\,\right )\,\right\rbrack =
\sqrt{\lambda \pi}\ \exp \left \lbrack\,\frac{1}{\lambda}\,\bigl (\kappa_{1}+\Ak_{2}\Ak_{3} \bigr)^{2}\,\right\rbrack\ ,
\end{equation}
\\
and (\ref{3.8}) is turned into

\begin{eqnarray}\label{3.12}
\Psi'(\vec\sigma)
& \propto &
\int\limits_{\Gamma_{3}} \dd\Ak_{3}\,  
\exp\left\lbrack\,\frac{1}{\lambda}\,\biggl(
\kappa_{1}^{2}+2 \Ak_{3}\kappa_{3}-\Ak_{3}^{2}\biggr)\,\right\rbrack\nonumber\\[0.5 cm]
& &\!\!\!\!\!\!
\mbox{\raisebox{0.3 ex}{$\cdot$ \hspace{-.5cm}}}
\int\limits_{\Gamma_{2}(\Ak_{3})} \!\!\dd\Ak_{2}\, 
\exp\left\lbrack\,\frac{1}{\lambda}\,\biggl(- \Ak_{2}^{2}\ (1-\Ak_{3}^{2})+2 \Ak_{2}\,(\kappa_{2}+\kappa_{1}\Ak_{3})\biggr)\,\right\rbrack\ .\ \  
\end{eqnarray}
\\
The  $\Ak_{2}$-integral is again of Gaussian form with just one saddle-point

\begin{equation}\label{3.13}
\Ak_{2}=\frac{\kappa_{2}+\kappa_{1}\Ak_{3}}{1-\Ak_{3}^{2}}\ .
\end{equation}
\\
To keep this saddle-point away from infinity, one has to exclude the points $\Ak_{3}=\pm 1$ from the integration path $\Gamma_{3}$. With this prescription and using the new coordinate

\begin{equation}\label{3.14}
\Ak_{2}'=\Ak_{2}\,\sqrt{1-\Ak_{3}^{2}}-\frac{\kappa_{2}+\kappa_{1}\Ak_{3}}{\sqrt{1-\Ak_{3}^{2}}}
\end{equation}
\\
the integration is again easily carried out along the real axis with the following result:

\begin{eqnarray}\label{3.15}
\Psi'(\vec\sigma)
&\propto &  
\int\limits_{\Gamma_{3}}\, \frac{\dd\Ak_{3}}{\sqrt{1-\Ak_{3}^{2}}} 
\ \exp\left\lbrack\,\frac{1}{\lambda}\,\biggl(
\kappa_{1}^{2}+2 \Ak_{3}\kappa_{3}-\Ak_{3}^{2}+\frac{(\kappa_{2}+\kappa_{1} \Ak_{3})^{2}}{1-\Ak_{3}^{2}}\,\biggr)\,\right\rbrack\nonumber\\[0.5 cm]
&=&  
\int\limits_{\Gamma_{3}} \frac{\dd\Ak_{3}}{\sqrt{1-\Ak_{3}^{2}}}\  
\exp\left\lbrack\,\frac{1}{\lambda}\,\biggl(\,
\frac{\kappa_{+}^{2}}{1-\Ak_{3}}+
\frac{\kappa_{-}^{2}}{1+\Ak_{3}}+2\, \Ak_{3}\, \kappa_{3}-\Ak_{3}^{2}\,\biggr)\,\right\rbrack\ . 
\end{eqnarray}
\\ 
Here further variables $\kappa_{\pm}:=\frac{1}{\sqrt{2}}\,(\kappa_{1}\pm \kappa_{2})$ have been introduced. Let us now define the new exponent as

\begin{equation}\label{3.16}
f(z,\vec \kappa):=\frac{\kappa_{+}^{2}}{1-z}+
\frac{\kappa_{-}^{2}}{1+z}+2\, z\, \kappa_{3}-z^{2} \ ,
\end{equation}
\\
then $\Psi'$ is easily expressed as

\begin{equation}\label{3.17}
\Psi'(\vec \sigma)\, \propto 
\int\limits_{\Gamma} \frac{\dd z}{\sqrt{1-z^{2}}}\ \exp\left\lbrack\, \frac{1}{\lambda}\,f(z,\vec \kappa)\,\right\rbrack\ .
\end{equation}
\\
Furthermore, a new coordinate $u=\mbox{Arcsin}\,z$ will prove useful, yielding the representation

\begin{equation}\label{3.18}
\Psi'(\vec \sigma)\, \propto 
\int\limits_{\CC} \dd u\ \exp\left\lbrack\, \frac{1}{\lambda}\,f(\sin u,\vec \kappa)\,\right\rbrack\ ,
\end{equation}
\\
where now the points $u=\pm \frac{\pi}{2}$ (and all $2 \pi$-periodic repetitions) have to be excluded from the new integration path $\CC$. It is possible to show that all saddle-points of $f$ are determined by the equation (\ref{3.7}) when $\Ak_{3}$ is replaced by $z$ and the $z$-solutions are translated to the $u$-plane afterwards. So the five saddle-points of the original integral representation (\ref{3.1}) occur in the one dimensional representation (\ref{3.18}), too, and we have still all freedom to choose a specific solution.

\begin{figure}
\begin{center}
\hskip 0 cm
\psfig{figure=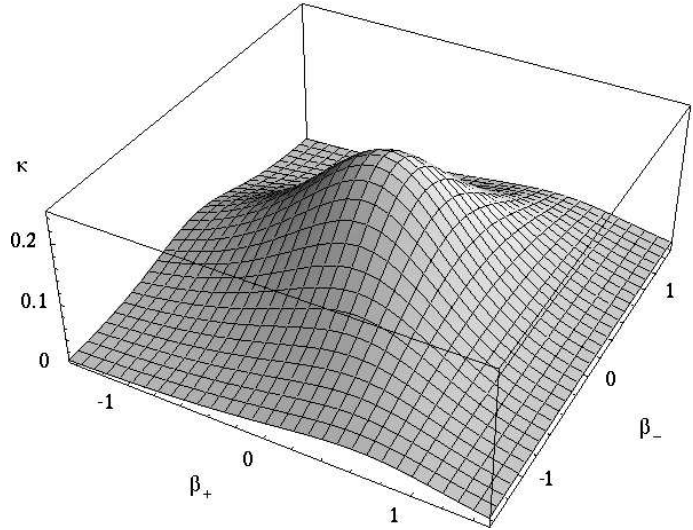,height=7cm}
\end{center}
\caption{The caustic in minisuperspace for $\Lambda > 0$}\label{grkaustik+}
\end{figure}

For a discussion of the location of the saddle-points we shall now restrict ourselves to the case $\Lambda >0$. Then it turns out that three of them are always real, which we will denote by

\begin{equation}\label{3.19}
z_{-}\leq -1\ \ ,\ \ -1 \leq z_{+} < 0\ \ ,\ \ z_{3}\geq 1
\end{equation}
\\
in the following. The other two solutions may also be real or conjugate complex depending on the values of the real positive variables $\kappa_{j}$. The two corresponding regions in the $\vec \kappa$-space (which is the minisuperspace with fixed $\Lambda$) are  separated by a {\em caustic} which is characterized by the existence of a {\em marginal} saddle-point of $f(z,\vec \kappa)$. We will refer to the first part of the minisuperspace, where all five saddle-points appear real in the $z$-plane, as the ``Euclidean regime'' and denote the two additional real saddle-points by\footnote{The naming of the indices introduced here will be justified later in the discussion of the asymptotic behaviour of the solutions.} 

\begin{equation}\label{3.20}
0 \leq z_{\Vil} \leq z_{0} \leq 1\ ,
\end{equation}
\\
whereas in the ``Lorentzian regime''  there exist two complex saddle-points labelled according to the signature of their imaginary parts as

\begin{equation}\label{3.21}
\Im\,z_{\Vil} < 0\ \Leftrightarrow \ \Im \, z_{\Vil}^{*} >0\ .
\end{equation}
\\
The caustic defined above will play an important role for some particular solutions which get their dominant integral contribution at the corresponding marginal saddle-point; for fig.$\!$ \ref{grkaustik+} it has been computed numerically in the $\{ \kappa, \bpm \}$-space (here $\kappa=\frac{1}{12}\, \Lambda a^{2}$, see below). If on the other hand solutions are considered which receive contributions from several distinct saddle-points, it should be clear that the caustic is of no significance to them. However, such solutions will turn out to be of little physical interest anyway. 

\begin{figure}
\begin{center}
\hskip 0 cm
\psfig{figure=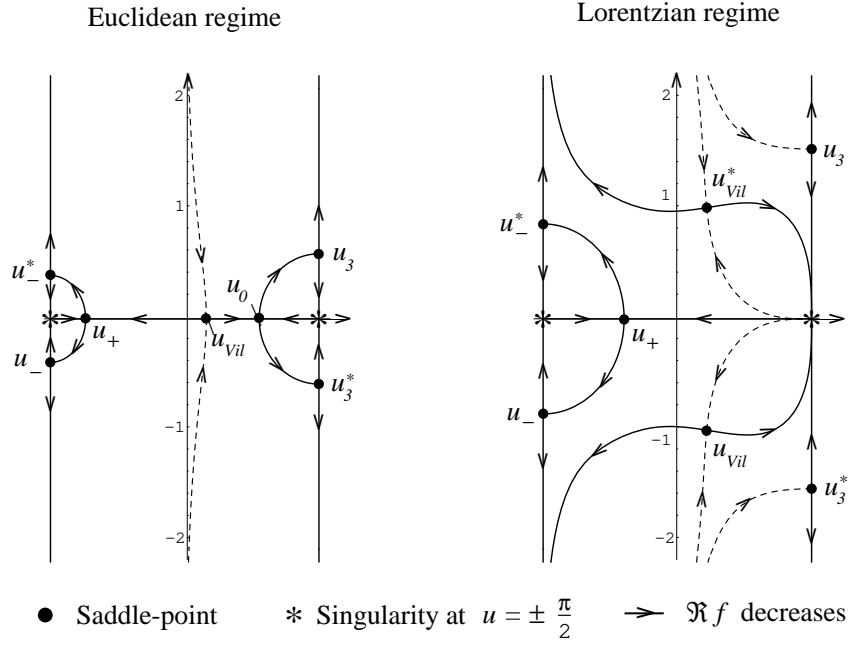,height=8.5cm}
\end{center}
\caption{Saddle-points and curves of steepest descent}\label{grcsd}
\end{figure}

\noindent
Knowing the saddle-points of $f$, one may now calculate the curves  of steepest descent for the two different regimes by solving the equations

\begin{equation}\label{3.22}
\Im\, f(\,\sin u(\tau)\, )=\Im \,f(\,\sin u_{\varrho}\,)\ \ ,\ \ \varrho\,\epsilon\,\{\mbox{{\footnotesize $-$}},\mbox{{\footnotesize $+$}},\mbox{{\scriptsize $Vil\,$}},0,3\}\ .
\end{equation}
\\
The result of a numerical approach to this problem is given in fig.$\!$ \ref{grcsd}. In the stripe $| \Re \, u |\leq \frac{\pi}{2}$ all solutions of (\ref{3.22}) are presented. The dashed curves reach to $+ \infty$ with respect to $\Re \, f$ and are given just for completeness. It is remarkable that there exist paths running into the singularities at $\Ak_{3}=\pm 1$, corresponding to $z = \pm 1$ and $u=\pm \frac{\pi}{2}$, in such a manner that $\Re \, f$ tends to $- \infty$. This results in additional possibilities to create integration contours obeying the requirements of the generalized Fourier-transformation. 

\vspace{.3 cm}

\subsection{Basis of linearly independent solutions}

With the knowledge of the curves of steepest descent now a basis of linearly independent solutions (\ref{3.1}) may be defined by choosing the following integration paths:

\vspace{.3 cm}
\begin{figure}
\begin{center}
\hskip 0 cm
\psfig{figure=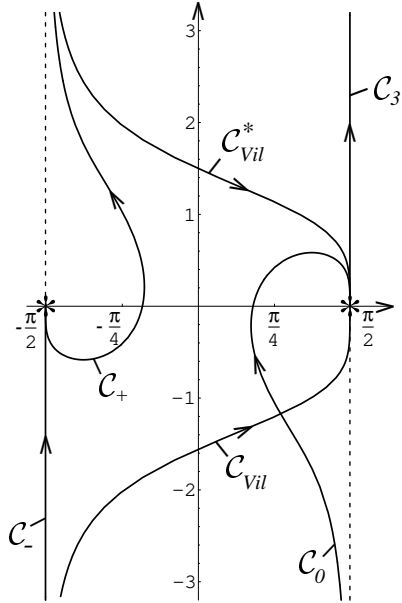,height=8cm}
\end{center}
\caption{Basis set of integration paths}\label{grintk}
\end{figure}

\noindent
The same paths can be chosen in the Euclidean and in the Lorentzian domain.
It is easily seen that all curves end at {\em essential} singularities of the integrand of (\ref{3.18}) in such directions that the integrand and {\em all} its $z$-derivatives vanish. This guarantees that 

\begin{equation}\label{3.23}
\Psi_{\varrho}'(\vec \sigma) \propto   
\int\limits_{\CC_{\mbox{{\scriptsize $\varrho$}}}} \dd u \ \exp\left\lbrack\, \frac{1}{\lambda}\,f(\sin u,\vec \kappa)\,\right\rbrack
\ , 
\ \ \varrho\,\epsilon\,\{\mbox{{\footnotesize $-$}},\mbox{{\footnotesize $+$}},\mbox{{\scriptsize $Vil\,$}},0,3\}
\end{equation}
\\
are indeed solutions of the Wheeler-DeWitt equation in the $\vec \sigma$-representation. Furthermore, {\em any} path satisfying the fall-off condition can be deformed into a superposition of the curves defined in figure \ref{grintk}, consequently one may express each wavefunction in terms of the corresponding solutions (\ref{3.23}). Thus the above mentioned basis property of (\ref{3.23}) is proven. 

For concreteness, let us write

\begin{equation}\label{3.24}
\Psi_{\varrho}(\vec \sigma)\,=\Psi^{0}_{\WH}\ {\cal N}_{\varrho}(\lambda)\ 
\int\limits_{\CC_{\mbox{{\scriptsize $\varrho$}}}} \dd u \ \exp\left\lbrack\, \frac{1}{\lambda}\,f(\sin u,\vec \kappa)\,\right\rbrack\ ,
\end{equation}
\\
where ${\cal N}_{\varrho}(\lambda)$ is a normalization factor that  usually is defined via
$\langle\Psi_{\varrho}|\Psi_{\varrho}\rangle
\ =^{^{\!\!\!\!\mbox{{\scriptsize !}}}}\ 1$. 
Here $\langle\cdot\, |\,\cdot\rangle$ denotes the scalar product of the underlying Hilbert space which we do not know explicitly. But even if we knew it, an analytical treatment of this normalization procedure would certainly not be tractable. That is why we adopt another convention and normalize the wavefunctions by the condition

\begin{equation}\label{3.25}
\Psi_{\varrho}\,(a=0)
\ \equiv^{^{\!\!\!\!\mbox{{\scriptsize !}}}}
\ 1\ ,
\end{equation}
\\
which implies that $\Psi_{\varrho}$ becomes independent of $\Lambda$ for $a \to 0$. This requirement gives analytically soluble normalization integrals, yielding

\begin{displaymath}
{\cal N}_{+}={\cal N}_{-}={\cal N}_{0}={\cal N}_{3}=-\frac{2 i \,e^{\mu}}{K_{0}(\mu)} \slambda  -\frac{2 i\,e^{\frac{1}{\lambda}}}{\sqrt{\lambda \pi}}\ ,
\end{displaymath}
\begin{equation}\label{3.26}
{\cal N}_{\Vil}=\frac{2 \,e^{\mu}}{2\pi I_{0}(\mu)+i K_{0}(\mu)}
\slambda \frac{1-\frac{i}{2}\,e^{-\frac{1}{\lambda}}}{\sqrt{\lambda \pi}}
\ \ ,\ \ \mbox{with}\ \ \mu:=\frac{1}{2 \lambda}\ ,
\end{equation}
\\
where $K_{0}$ and $I_{0}$ are the McDonald's and the modified Bessel function with index $0$, respectively. The asymptotic behavior of ${\cal N}_{\varrho}$ in the limit $\lambda \to 0$ will be useful in the next section.

There are other solutions of the Wheeler-DeWitt equation which will turn out to be rather interesting, and which of course can be expressed in terms of the basis solutions (\ref{3.24}):
By choosing another order of integrations in (\ref{3.8}) the wavefunctions $\Psi_{1,2}$ defined by  

\begin{equation}\label{3.27}
\Psi_{1}(\kappa_{1},\kappa_{2},\kappa_{3}):=
\Psi_{3}(\kappa_{2},\kappa_{3},\kappa_{1}) 
\ \ \mbox{and}\ \  
\Psi_{2}(\kappa_{1},\kappa_{2},\kappa_{3}):=
\Psi_{3}(\kappa_{3},\kappa_{1},\kappa_{2})
\end{equation}
\\
apparently solve the Wheeler-DeWitt equation, and the asymptotic behavior in section \ref{4} will show that the relations

\begin{equation}\label{3.28}
\Psi_{1}=\biggl\{
\begin{array}{ccc}
\Psi_{+}\!\!&,&\kappa_{1} \geq \kappa_{2}\\
\Psi_{-}\!\!&,&\kappa_{1} \leq \kappa_{2}
\end{array}\qquad ,\qquad 
\Psi_{2}=\biggl\{
\begin{array}{ccc}
\Psi_{+}\!\!&,&\kappa_{1} \leq \kappa_{2}\\
\Psi_{-}\!\!&,&\kappa_{1} \geq \kappa_{2}
\end{array}\ 
\end{equation}
\\
hold in the limit $\Lambda\, a^{2} \to \infty$. Since (\ref{3.24}) is a {\em basis} with non-vanishing $\Psi_{\varrho}$ in this limit, it follows that (\ref{3.28}) must hold in general. This claim is proven by considering a general expansion in terms of the basis states, with coefficients depending on $\Lambda$, but not on $a$, where one makes use of the limit $a \to \infty$ at fixed $\Lambda \not= 0$. 

With the solutions $\Psi_{1}$ and $\Psi_{2}$, now the sum

\begin{equation}\label{3.29}
\Psi_{+}+\Psi_{-}+\Psi_{3} \equiv \Psi_{1}+\Psi_{2}+\Psi_{3}
\end{equation}
\\
turns out to be a solution as well, which is {\em symmetric} with respect to arbitrary permutations of the $\kappa_{j}$. If in addition one uses the fact that the manifolds

\begin{equation}\label{3.30}
\Sigma_{\pm}:=\left \{\  \vec \Ak \,\epsilon\, \C^{3}\  |
\ \Ak_{j}\, \epsilon \,\R \,e^{\pm i \frac{\pi}{6}},\,j\, \epsilon\,\{1,2,3\}
\ \right \}\ ,
\end{equation}
\\
which are symmetric under permutations of the $\Ak_{j}$, are suitable to perform the generalized Fourier-transformation in the representation (\ref{3.1}), sucessive integrations analogous to (\ref{3.8})-(\ref{3.15}) reveal that $\Psi_{0}$ and $\Psi_{\Vil}$ are symmetric under permutations of the $\kappa_{j}$.

Finally, by examining the integrand's symmetries under complex conjugation, $\Psi_{\varrho}$ turns out to be a real wavefunction for $\varrho\, \epsilon\, \{\mbox{{\footnotesize $-$}},\mbox{{\footnotesize $+$}},0,1,2,3\}$ , whereas $\Psi_{\Vil}$ is complex-valued in general.

\section{Asymptotic results}\label{4}

In this section we will discuss the asymptotic behavior of the solutions to the Wheeler-DeWitt equation with non-vanishing cosmological constant derived above. This will be done by evaluating the integral representation (\ref{3.24}) in the limits $\hbar \to 0$ and $\Lambda \to 0$, where (\ref{3.24}) turns out to assume the form of a saddle-point integral. Consequently, the asymptotically leading term of such an integral

\begin{equation}\label{4.1}
\int \limits_{\CC} \dd z \, e^{\frac{1}{\lambda}\,f(z)} \slambda 
\pm \sqrt{- \frac{2 \pi \lambda}{f''(z_{s})}}\,  e^{\frac{1}{\lambda}\,f(z_{s})}
\end{equation}
\\
will be of particular interest \cite{25,26}. Here $z_{s}$ denotes that saddle-point of $f$ which provides the only contribution to the integral in the limit $\lambda \to 0$, and the sign of the square root has to be adjusted to the direction in which this saddle-point is passed through. In the following ``$\sqrt{\ \ }\,$'' will denote the principal value of the square-root.

The application of (\ref{4.1}) to the general representation (\ref{3.24}) yields 

\begin{eqnarray}\label{4.2}
\Psi_{\varrho}(a,\beta_{\pm})& \slambda &\Psi^{0}_{\WH}\ {\cal N}_{\varrho}(\lambda)\ \sqrt{ \pi \lambda}\  \left \lbrack\ 
\frac{\exp \Bigl \lbrack\frac{1}{\lambda} \ f(\sin u) \Bigr \rbrack}{\sqrt{-\frac{1}{2}\,\frac{\mbox{d}^{2}}{\mbox{d$u$}^{2}} f(\sin u)}}\ \right \rbrack_{\mbox{{\footnotesize $u=u_{\varrho}$}}}
\nonumber\\[0.5 cm]
&=&{\cal N}_{\varrho}(\lambda)\ \sqrt{ \pi \lambda}\ 
\left\lbrack \ \frac{\exp\Bigl\lbrack \frac{1}{\hbar}\, \Bigl(
-\Phi +\frac{6 \pi}{\Lambda}\, f(z)\,
\Bigr) \Bigr \rbrack}
{\sqrt{\frac{1}{2}\,\Bigl(f''(z)(z^{2}-1)+f'(z)\,z\Bigr)}}\ \right\rbrack_{\mbox{{\footnotesize $z=z_{\varrho}$}}} \ \  ,
\ \ \varrho \, \epsilon \, \{\mbox{{\footnotesize $-$}},\mbox{{\footnotesize $+$}},\mbox{{\scriptsize $Vil\,$}},3\}
 \ .
\end{eqnarray}
\\
Since $\lambda=\frac{\hbar \Lambda}{6 \pi}$, this formula includes both, the limit $\hbar \to 0$ and $\Lambda \to 0$.
That the curves defined in fig.$\!$ \ref{grintk} give indeed the saddle-point contributions mentioned in (\ref{4.2}) can be extracted from fig.$\!$ \ref{grcsd} by analyzing the topological properties of $f$ in detail. The asymptotic expansion (\ref{4.2}) also holds for $\Psi_{0}$, as long as just the Euclidean regime is considered. In the Lorentzian case {\em all} saddle-points can be passed through by $\CC_{0}$ and may therefore contribute, so in dependence on the variables $\alpha$ and $\bpm$ one has to choose the highest saddle-point to employ (\ref{4.2}).

The result (\ref{4.2}) now suggests the definitions 

\begin{eqnarray}\label{4.3}
&S_{\varrho}:=i\, \Phi -\frac{6 \pi i}{\Lambda} 
\,f(z_{\varrho}) \ \ ,\ \  
A_{\varrho}:=\Biggl \lbrack\,\frac{1}{2}\Bigl ( f''(z_{\varrho})(z_{\varrho}^{2}-1)+f'(z_{\varrho})\,z_{\varrho}\,\Bigr )\, \Biggr \rbrack^{-\frac{1}{2}}&\nonumber\\[0.5 cm]
&\Rightarrow \qquad \Psi_{\varrho} \slambda \sqrt{\pi \lambda}\ {\cal N}_{\varrho}\,A_{\varrho}\,\exp \Bigl \lbrack \frac{i}{\hbar}\,S_{\varrho} \Bigr \rbrack \ .&
\end{eqnarray}
\\
In the limit $\hbar \to 0$ the exponent $S$ can then be interpreted directly as the action of the wavefunction (up to a constant term, which may arise from ${\cal N}_{\varrho}$), while $A$ plays the role of Gaussian fluctuations around the saddle-point $z_{\varrho}$. Considering the limit $\Lambda \to 0$, these interpretations hold no longer; here $S$ and $A$ may just be called the phase- and amplitude function of $\Psi$, respectively.

Unfortunately, equation (\ref{3.7}) determining the saddle-points is of fifth order, so analytical expressions for the roots are not available. This is why we shall first restrict ourselves to the isotropic case $\bpm=0$, where (\ref{3.7}) can be solved explicitly, yielding 

\begin{equation}\label{4.4}
z_{\pm}^{(0)}=-1\ ,\ \ z_{3}^{(0)}=1+\kappa\ ,\ \ z_{\Vil}^{(0)}=\frac{1}{2}-\sqrt{\,\frac{1}{4}-\kappa}\ , \ \ 
\frac{1}{2}+\sqrt{\,\frac{1}{4}-\kappa}=\left \lbrace
\begin{array}{ccc}
z_{0}^{(0)}&,& \kappa \leq \frac{1}{4}\\[0.4 cm]
z_{\Vil}^{(0)\,*}&,& \kappa \geq \frac{1}{4}
\end{array}
\right.\ ,\ \mbox{where}\ \ \kappa:=\frac{1}{12}\, \Lambda a^{2}\ . 
\end{equation}
\\
With these preparations we shall now turn to a detailed discussion of the specific solutions.

\subsection{Vilenkin state}

\subsubsection{Semi-classical limit $\hbar \to 0$}

\noindent
If the saddle-point $z^{(0)}_{\Vil}$ is inserted into (\ref{4.3}) one obtains the action

\begin{equation}\label{4.5}
S_{\Vil}^{(0)}=\frac{3\, i \pi}{\Lambda} \, \left \lbrack\,
1-\left( 1-\frac{1}{3} \, a^{2} \, \Lambda   \right)^{\frac{3}{2}} \,\right \rbrack
\end{equation}
\\
for the isotropic case. Choosing the lapse function $N \equiv 1$, the only non-trivial equation for the classical trajectories reads

\begin{equation}\label{4.6}
\frac {\dd a}{\dd t}=-\frac{1}{3 a \pi}\,\frac{ \partial S_{\Vil}^{(0)}}{\partial a}
=- i\, \sqrt{\ 1-\frac{1}{3}\,a^2\,\Lambda}
\end{equation}
\\
and is easily integrated to give 

\begin{equation}\label{4.7}
a(t)=a_{0}\,
\cosh \left (a_{0}^{-1}\,t \right ) \ ,\ \ \mbox{where}\ \ a_{0}:=\sqrt{\frac{3}{\Lambda}}\ ,\ \ t >0\ .
\end{equation}
\\
Reinserting this classical {\em DeSitter} solution into (\ref{4.6}) and remembering that ``$\sqrt{\ \ }\,$'' denotes the principal value of the square-root reveals that, as mentioned, (\ref{4.6}) is solved only with the restriction to $t > 0$, i.e. collapsing Universes are not described by $\Psi_{\Vil}$.    

Metrics with $a < a_{0}$ can be obtained by solving the Euclidean version of (\ref{4.6}) with $\dd \tau = i \,\dd t$, yielding 

\begin{equation}\label{4.8}
a(\tau)=a_{0}\,
\sin \left (a_{0}^{-1}\,\tau \right )\ ,\ \ \frac{a_{0}\,\pi}{2}
\leq \tau < a_{0} \pi\ .
\end{equation}
\\
The restriction of the $\tau$-variable appears for the same reason as discussed above for (\ref{4.7}). Denoting the line element of the unit 3-sphere by $\dd \Omega$, the 4-metric corresponding to (\ref{4.8}) reads

\begin{equation}\label{4.9}
\dd s^{2}=a_{0}^{2}\,\left (\, \dd \tau'^{2}+
\sin^{2} \tau'\, \dd \Omega^{2}\, \right ) \ ,\ \ \mbox{with}\ \ \tau' := 
a_{0}^{-1}\,\tau\ .
\end{equation}
\\
It describes exactly a 4-half-sphere with radius $a_{0}$, which in the limit $\tau' \to \frac{\pi}{2}$ may be extended to the DeSitter solution (\ref{4.7}). Furthermore, the point $a=0 \Leftrightarrow \tau' = \pi$ is a regular point of the manifold, i.e. $\Psi_{\Vil}$ satisfies the no-boundary proposal {\em in the isotropic case}.  

It is interesting to see whether these properties remain true when anisotropic corrections are considered. The calculation of such corrections is straightforward: One has to expand $z_{\Vil}$ for small $\bpm$, and insert this expansion in the expression for $S_{\Vil}$. One finally obtains\footnote{A corresponding expression for $S_{\Vil}$, which is here obtained as a limit of the exact result, was first derived by Del Campo and Vilenkin, using the WKB method \cite{21}. These authors did {\em not} discuss the consequences for the semi-classical trajectories generated by this action.} 

\begin{equation}\label{4.11}
S_{\Vil}=S_{\Vil}^{(0)}+36 \pi  i\,a^2\,\frac{3+\sqrt{\,1-\frac{1}{3}\,a^{2} \Lambda}}
{24+a^{2} \Lambda}\, (\beta_{+}^{\, 2}+\beta_{-}^{\,2})+{\cal O}(\beta_{\pm}^{\, 3})\ .
\end{equation}
\\
This action yields for the classical trajectories in imaginary time in leading order of $\bpm$ (with $\kappa=\frac{1}{12}\, \Lambda a^{2}$) 

\begin{equation}\label{4.12}
\frac{\dd a}{\dd \tau} \approx  
-\sqrt{\ 1-4\kappa}+4\ \frac{\beta_{+}^{\,2}+\beta_{-}^{\,2}}{(\kappa+2)^{2}}\, 
\left \lbrack\, 
\frac{(\kappa+3)^2-10}{\sqrt{\ 1- 4 \kappa}} -3\,\right \rbrack
\  ,
\end{equation}

\vspace{0.2 cm}

\begin{equation}\label{4.13}
\frac{\dd \bpm}{\dd \tau}
\approx \frac{2}{a}\  \frac{3+\sqrt{\ 1- 4 \kappa}}{2+\kappa}\, \bpm\ ,
\end{equation}
\\
where we have assumed $\kappa < \frac{1}{4}$. Since the prefactor of $\bpm$ in (\ref{4.13}) is positive definite, the anisotropy decreases with decreasing $\tau$, while (\ref{4.12}) then tells us that $a$ tends to $a_{0}$. Consequently, a flat, cylindrical 4-geometry

\begin{equation}\label{4.14}
\dd s^{2} = \dd \tau^{2}+a_{0}^{2} \, \dd \Omega^{2}\, \geq 0
\end{equation}
\\
is approached in this limit with $a_{0}$ playing the role of the cylinder radius.

With increasing $\tau$ the point $a=0$ is approached and the anisotropy grows rapidly so that the validity of (\ref{4.12}) and (\ref{4.13}) breaks down. Therefore $a=0$ is not a regular point of the Euclidean space-time manifold any more, and the no-boundary proposal is {\em not} fulfilled for $\Psi_{\Vil}$.

A discussion of a Lorentzian version of (\ref{4.12}) and (\ref{4.13}) reveals that classical DeSitter-like Universes are described, which grow exponentially in time $t$ while the anisotropy decreases monotonously.\footnote{In the Lorentzian regime the action calculated from (\ref{4.3}) has a non-vanishing real- {\em and} imaginary part for $\bpm \not= 0$, so here the definition of classical trajectories is not a priori clear. Since we are interested in pseudo-Riemannian 4-geometries, and because $\Re S_{\Vil}$ dominates $\Im S_{\Vil}$ for large scale parameters $a$ anyway, we choose the {\em real part} of $S_{\Vil}$ to discuss the Lorentzian classical trajectories.} However, in general $\bpm$ does not tend to zero, but approaches a finite value. 

In any case $\Psi_{\Vil}$ describes an {\em expanding} Universe, i.e. quantum mechanically speaking $\Psi_{\Vil}$ supports a current in minisuperspace which is directed to the positive $a$-axis. Thus this wavefunction satisfies the condition proposed by {\em Vilenkin} and so is identified as the {\em Vilenkin state} of the Bianchi type IX model with cosmological constant.

\subsubsection{The limit $\Lambda \to 0$}

In contrast to the limit $\hbar \to 0$, in the case $\Lambda \to 0$ the location of the saddle-points themselves depends on $\Lambda$ via the variables $\kappa_{j}$. Nevertheless, the expansion (\ref{4.2}) remains applicable as long as just the leading term is considered, and happily now the $\Lambda$-corrections of the saddle-points can be calculated taking account of the full influence of anisotropy. The first terms read

\begin{equation}\label{4.15}
z_{\Vil}=\frac{1}{2}\,\sigma_{3}\, \lambda+\frac{1}{4}\,\sigma_{1} \sigma_{2}\, \lambda^{2}+
{\cal O}(\lambda^{3})\ ,
\end{equation}
\\
and the phase- and amplitude functions are calculated to be

\begin{eqnarray}\label{4.16}
&S_{\Vil}=i\, \Phi-
i \,\hbar \,\frac{\lambda}{4}\,
(\sigma_{1}^{2}+ \sigma_{2}^{2}+\sigma_{3}^{2}) \, 
+{\cal O}(\Lambda^{2})\ ,&\nonumber\\[0.4 cm]
&A_{\Vil}=1+\frac{\lambda^{2}}{8}\,
\,(\sigma_{1}^{2}+ \sigma_{2}^{2}+\sigma_{3}^{2})+
{\cal O}(\Lambda^{3})\ .&
\end{eqnarray}
\\
Using (\ref{4.3}) and the asymptotic behavior of ${\cal N}_{\varrho}$ in accordance with (\ref{3.26}), one finds for the wavefunction:

\begin{equation}\label{4.17}
\Psi_{\Vil}
\slambda
{\cal N}_{\Vil}\,\sqrt{\pi \lambda}\  A_{\Vil}
\  e^{\mbox{{\footnotesize $\frac{i}{\hbar}\,S_{\Vil}$}}}
\tlambda \Psi_{\WH}^{0}
\ .
\end{equation}
\\
Thus in the limit $\Lambda \to 0$ the wormhole state of the $\Lambda=0$ -model is approached.

\subsubsection{The limit $\kappa \to \infty$}

\noindent
Writing the wavefunctions $\Psi_{\varrho}$ defined in (\ref{3.24}) in the alternative form

\begin{equation}\label{4.18}
\Psi_{\varrho}=\Psi_{\WH}^{0}\,{\cal N}_{\varrho}
\,\int\limits_{\CC_{\varrho}} \dd u\, 
\exp \Bigg\lbrack\, 
\frac{12 \pi\, \kappa^{2}}{\hbar \Lambda} \,
\Bigg (  \, \frac{\sin u+\cosh \left \lbrack 
2 \sqrt{3} \,\beta_{-}\right \rbrack}
{\cos^{2} u}\,e^{-2 \beta_{+}}-\frac{1}{2}\, \left(\frac{\sin u}{\kappa} \right )^{2}+\frac{\sin u}{\kappa} \,e^{2 \beta_{+}} \,
\Bigg ) \Bigg \rbrack\ ,
\end{equation}
\\
another possibility to approximate a saddle-point integral occurs, namely the limit $\kappa=\frac{1}{12}\,\Lambda a^{2} \to \infty$. This includes the cases $\Lambda \to \infty$ and, in particular, $a \to \infty$.

An asymptotic expansion of the saddle-points is again possible without any restriction for the anisotropy variables, but the result is rather lengthy. So we turn at once to the expressions for $S_{\Vil}$ and $A_{\Vil}$ obtained in this limit:

\begin{equation}\label{4.19}
S_{\Vil}=
\frac{6 \pi}{\Lambda}\, \Bigg \{\!
-4 \,\sqrt{\kappa}^{\,3}+\, 
\sqrt{\kappa}\ \,\mbox{Tr} \left ( e^{-2 \mbox{{\footnotesize $\beta$}}}-\frac{1}{2}\,
e^{4 \mbox{{\footnotesize $\beta$}}}
 \right )+\frac{i}{2}\,\Biggl \lbrack\,
7-
\mbox{Tr}\, e^{2 \mbox{{\footnotesize $\beta$}}}\!  \cdot
\mbox{Tr}\,  e^{-2 \mbox{{\footnotesize $\beta$}}} 
+\mbox{Tr}\,  e^{6 \mbox{{\footnotesize $\beta$}}} \,
\Biggr \rbrack \,\Bigg \}
+{\cal O}(\kappa^{-\frac{1}{2}})\ ,  
\end{equation}

\vspace{0.2 cm}

\begin{equation}\label{4.20} 
A_{\Vil}=\frac{1}{\sqrt{2}}\,
e^{\frac{i\, \pi}{4}}\,
 \kappa^{-\frac{3}{4}}\, \left \{
1+\frac{i}{2}\,\frac{1}{\sqrt{\kappa}}\,
\mbox{Tr} \, e^{2 \mbox{{\footnotesize $\beta$}}}  \right \}+
{\cal O}(\kappa^{-\frac{7}{4}})\ . 
\end{equation}
\\
Surprisingly, the contribution $i \Phi$, which usually arises from the wormhole state and would be expected to give a term of ${\cal O}(\kappa)$ in (\ref{4.19}) has completely disappeared.

To shorten the final expression for $\Psi_{\Vil}$, let us now expand for small mean anisotropy $\beta :=\sqrt{\beta_{+}^{\,2}+\beta_{-}^{\,2}}$ :

\begin{eqnarray}\label{4.21}
\Psi_{\Vil}\ 
&\skappabeta&
\frac{\sqrt{\hbar}}{K_{0}\left(\frac{3 \pi}{\hbar \Lambda} \right)
-2 i \pi I_{0}\left (\frac{3 \pi}{\hbar \Lambda} \right )}\,
\left (\frac{3}{\Lambda} \right )^{\frac{1}{4}}\,
\biggl ( \frac{a}{2} \biggr )^{-\frac{3}{2}}\  \Bigg \{
1+ \frac{i}{a}\, \sqrt {\frac{3}{\Lambda}}\, \left (
3+12 \beta^{2} \right ) \Bigg \}
\nonumber\\[0.6 cm]
& &\qquad \cdot \ e^{^{\mbox{{\footnotesize$
-\frac{3 \pi}{\hbar \Lambda}\, (6 \beta)^{2}$}}}} \exp \left \lbrack \frac{\pi i}{\hbar}\, \sqrt{
\frac{\Lambda}{3}} \, \Bigg (
-a^{3}+\frac{a}{\Lambda}\, \left ( \frac{9}{2}-36\, \beta^{2} \right ) \Bigg ) -\frac{i \pi}{4}\,\right \rbrack\ .  
\end{eqnarray}
\\
From this result it is clear that $|\Psi_{\Vil}|^{2}$ is bounded for $a \to \infty$. Furthermore, a saddle-point expansion for $\beta \to \infty$ at fixed $a$ reveals that $\Psi_{\Vil}$ is square-integrable over $\bpm$, and because it is bounded for $\alpha \to -\infty$ (cf. (\ref{3.25})) it is normalizable in the distributional sense.\footnote{This entails the possibility to use Marolf's method \cite{27} for introducing a scalar product to define a Hilbert space of physical states, which contains $\Psi_{\Vil}$, presumably as a ground state.} 

As a general result the Vilenkin state becomes concentrated about $\bpm=0$ in the limit $\kappa \to \infty$, but with a non-vanishing Gaussian width

\begin{equation}\label{4.21+}
\Delta \beta \skappa \frac{1}{6}\, \sqrt{\frac{\hbar \Lambda}{6 \pi}}\ .
\end{equation}
\\
To give an idea of the behavior of the exact analytical solution we have computed the real- and imaginary part of the wavefunction numerically in dependence on $a$ and $\Lambda$, assuming $\bpm=0$ and picking units with $\hbar= 2 \pi$ :\footnote{This unusual choice of $\hbar$ is due to the use of differently scaled variables in the numerical work and is of course of little significance to the figure.}

\vspace{.3 cm}
\begin{figure}
\hskip 0 cm
\psfig{figure=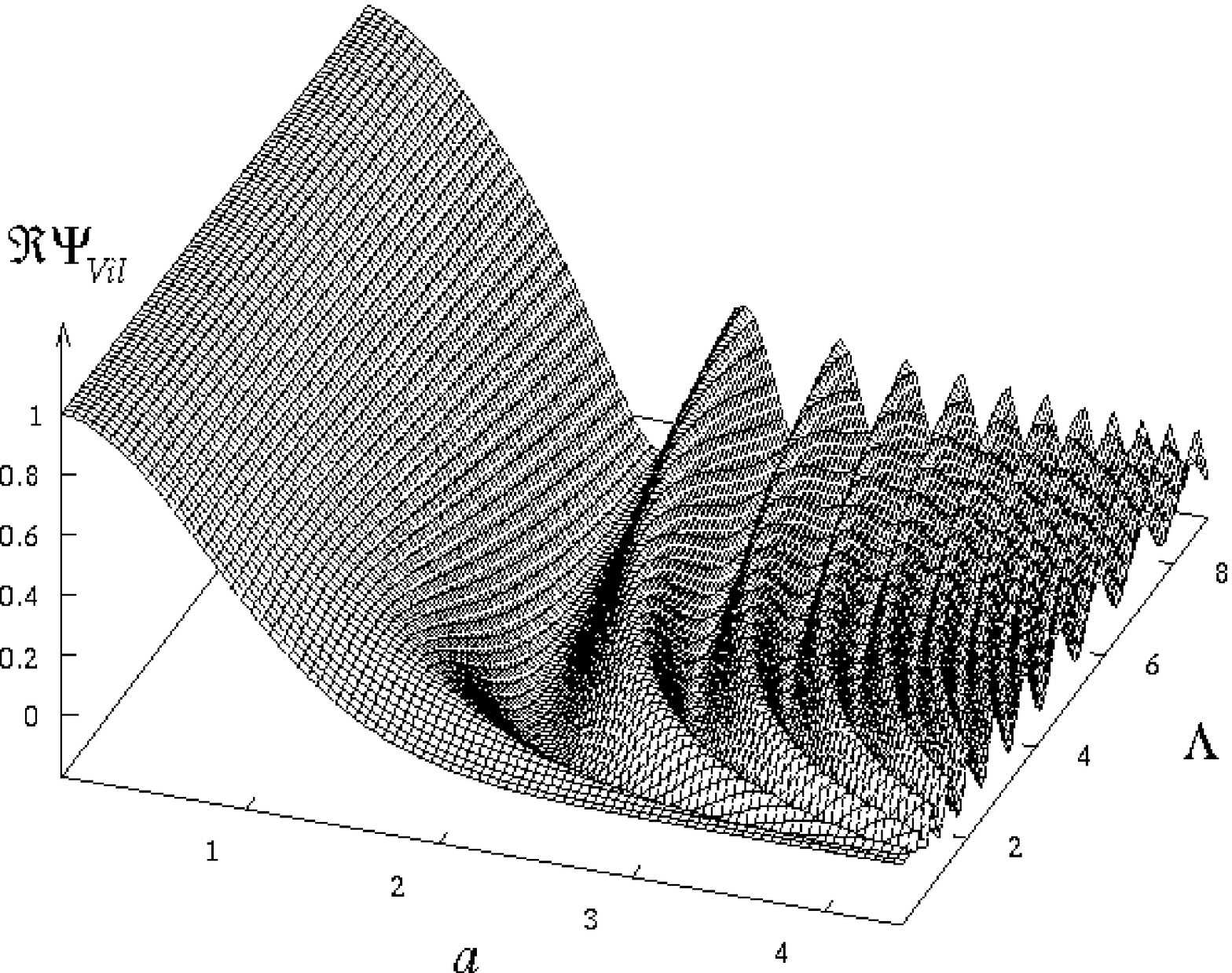,width=8cm}
\hskip 1 cm
\psfig{figure=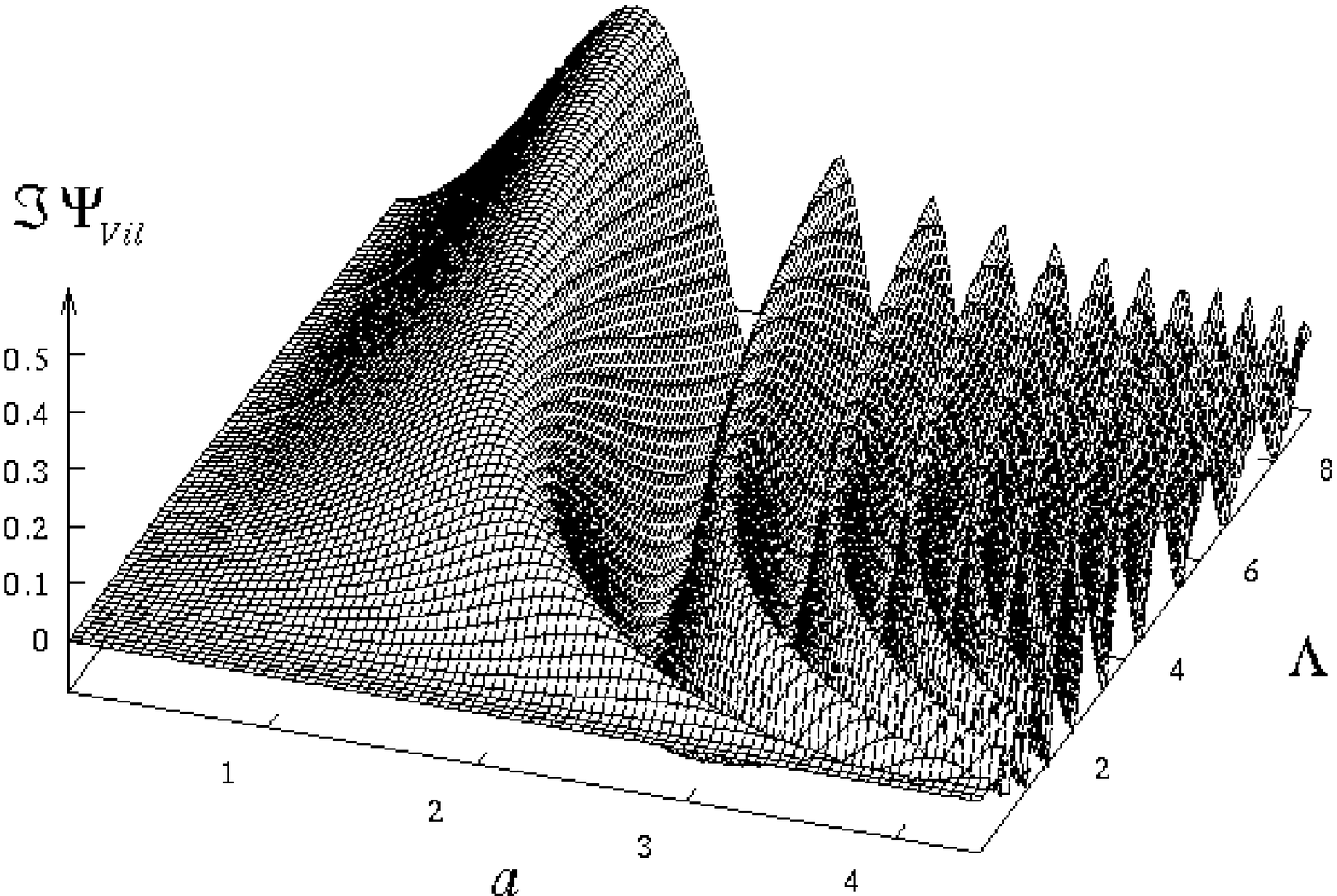,width=8cm}
\caption{The Vilenkin state}\label{gravil}
\end{figure}
\vspace{.3 cm}

\noindent
The conspicious oscillations, which start on the caustic $\kappa= \frac{1}{4} \Leftrightarrow \Lambda a^{2}=3$, indicate that the initial Euclidean action has turned Lorentzian, corresponding to a pseudo-Riemannian classical Universe. For $\Lambda \to 0$ the Gauss function of the wormhole state is recovered, whereas our normalization condition (\ref{3.25})
is responsible for the form of the graph at $a=0$.

\begin{figure}
\begin{center}
\hskip 0 cm
\psfig{figure=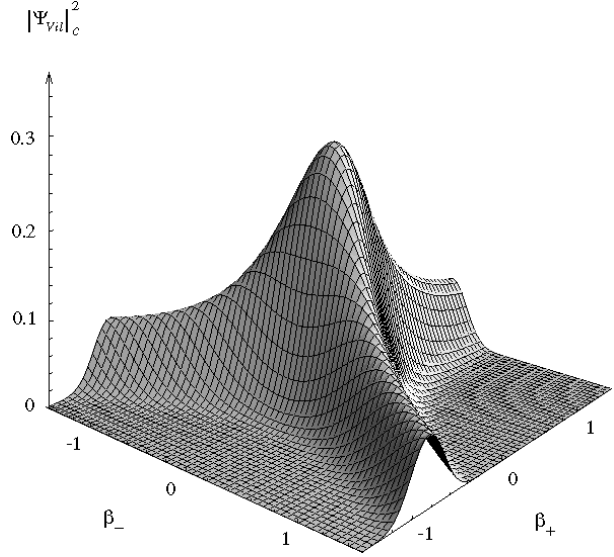,width=8cm}
\end{center}
\caption{The initial value distribution generated by the Vilenkin state}\label{grkv}
\end{figure}

In addition, the value of $|\Psi_{\Vil}|^{2}_{c}$  {\em on the caustic} has been computed for fig.$\!$ \ref{grkv} in dependence on the anisotropy variables $\bpm$ at fixed values $\Lambda=3, \,\hbar=2 \pi$. Since all the Lorentzian trajectories end at the caustic, we follow Hawking and suggest to interpret $|\Psi|^{2}_{c}$ as the distribution of the initial values for a classical evolution of the Universe. For the Vilenkin wavefunction the distribution is nicely concentrated about $\bpm=0$, but there are directions of the $\bpm-$plane in which $|\Psi_{\Vil}|^{2}_{c}$ takes a {\em finite} value for $\beta \to \infty$. Moreover, one may show that in these directions a tube with finite height and {\em width} is approached. This is not in contradiction to the normalizability of the wavefunction in the distributional sense, since on the caustic $\alpha \to -\infty$ in the same limit. However, it implies that $|\Psi_{\Vil}|^{2}_{c}$ is not square-integrable with respect to the $\bpm$-variables!

\hspace{.3 cm}

\noindent
In subsection C we will define a new state as a superposition of $\Psi_{\Vil}$ and $\Psi_{\Vil}^{*}$ in such a way, that this undesirable property of the initial value distribution is removed.

\subsection{Asymmetric states}

\subsubsection{Semi-classical limit $\hbar \to 0$}

In the isotropic case $\bpm =0$ the actions of the wavefunctions $\Psi_{\pm}$ and $\Psi_{3}$ are easily calculated as the corresponding saddle-point contributions in accordance with (\ref{4.4}):

\begin{equation}\label{4.22}
S_{\pm}^{(0)}=S_{3}^{(0)}=\frac{6 \pi i}{\Lambda}\,
\left \lbrack 1+5 \left ( \frac{a}{2\,a_{0}} \right )^{2}
\!-\left ( \frac{a}{2\,a_{0}} \right )^{4} \right \rbrack \ .
\end{equation}
\\
Obviously these actions are purely Euclidean for generic $a$ and $\Lambda$, and this statement remains true if anisotropic corrections are considered. An additional discussion of the limit $\kappa \to \infty$ will show that the asymmetric solutions themselves are indeed of no physical relevance, nevertheless, an examination of the case $\Lambda \to 0$ will be worthwile.

\subsubsection{The limit $\Lambda \to 0$}

As in the Vilenkin case, in the limit $\Lambda \to 0$ asymptotic expressions for the saddle-points $z_{\pm}$ and $z_{3}$ are available taking full account of anisotropy. The results read

\begin{equation}\label{4.23}
z_{\pm}=-1 \pm \frac{ \lambda}{4}\ |\sigma_{1}-\sigma_{2}| +{\cal O}(\lambda^{2})\ ,\qquad 
z_{3}=1 + \frac{ \lambda}{4}\,
(\sigma_{1}+\sigma_{2}) 
+{\cal O}(\lambda^{2})\ ,
\end{equation}
\\
and using them, the calculation of the phase- and amplitude functions is straightforward, yielding

\begin{equation}\label{4.24}
S_{\pm}= i\,\biggl \lbrack\, \frac{6 \pi}{\Lambda}+\Phi + \hbar\,(\,\sigma_{3} \mp |\sigma_{1}-\sigma_{2}|) \,\biggr \rbrack+{\cal O}(\Lambda)\ ,\qquad
S_{3}= i\,\biggl \lbrack\,\frac{6 \pi}{\Lambda}+ \Phi - \hbar\,(\,\sigma_{3} -\, \sigma_{1}-\,\sigma_{2}\,) \,\biggr \rbrack+{\cal O}(\Lambda)\ ,
\end{equation}
and
\begin{equation}\label{4.25}
A_{\pm}=\frac{i}{2}\,\biggl \lbrack\, 1-\frac{ \lambda}{4}(\,\sigma_{3} \mp |\sigma_{1}-\sigma_{2}| )\, \biggr \rbrack+{\cal O}(\Lambda^{2})\ ,\qquad
A_{3}=\frac{i}{2}\,\biggl \lbrack\, 1 +\frac{ \lambda}{4}(\,\sigma_{3}-\,\sigma_{1}-\,\sigma_{2}\,) \,\biggr \rbrack+{\cal O}(\Lambda^{2})\ .
\end{equation}
\\
Thus one obtains for the wavefunctions in the limit $\Lambda \to 0$

\begin{equation}\label{4.26}
\lim_{\Lambda \to 0}\, \Psi_{\pm} =
\Psi_{\WH}^{0} \,  \exp \Bigl \lbrack 
 -\sigma_{3} \pm \, | \sigma_{1}-\sigma_{2}| \Bigr \rbrack ,\qquad
\lim_{\Lambda \to 0}\, \Psi_{3} =
\Psi_{\WH}^{0}\,\exp \Bigl \lbrack \sigma_{3}-\sigma_{1}-\sigma_{2} \Bigr \rbrack =\Psi_{3}^{0}\ ,
\end{equation}
\\
where, in addition, the asymptotic behavior (\ref{3.26}) of the normalization factors has been taken into account. Obviously the solutions $\Psi_{\pm}$ are not differentiable at $\sigma_{1}=\sigma_{2}$, i.e. in particular at $\beta_{-}=0$, but if they are replaced by $\Psi_{1, 2}$ via (\ref{3.28}), solutions with nice analytical properties are obtained. These are related to $\Psi_{3}$ by permutations of the $\sigma_{j}$ (or equivalently the $\kappa_{j}$) as mentioned in (\ref{3.27}).

The wavefunctions $\Psi_{i}, \,i \, \epsilon \, \{ 1, 2, 3 \}$ are now easily recognized as {\em asymmetric} states, which approximate the corresponding solutions $\Psi_{i}^{0}$ of the $\Lambda=0$ -model when $\Lambda$ tends to zero.

\subsubsection{The limit $\kappa \to \infty$}

\noindent
Let us first consider the case $\kappa_{1} > \kappa_{2}$, where the saddle-points can be expanded as follows:

\begin{equation}\label{4.27}
z_{-}=-\frac{\kappa_{1}}{\kappa_{2}}+
\frac{\kappa_{3}\,(\kappa_{1}^{2}-\kappa_{2}^{2})}{\kappa_{2}^{4}}+{\cal O}(\kappa^{-2})\ , \ \ 
z_{+}=-\frac{\kappa_{2}}{\kappa_{1}}+
\frac{\kappa_{3}\,(\kappa_{2}^{2}-\kappa_{1}^{2})}{\kappa_{1}^{4}}+{\cal O}(\kappa^{-2})\ , \ \  
z_{3}\,= \kappa_{3}+\frac{\kappa_{1}\,\kappa_{2}}{\kappa_{3}^{2}}+{\cal O}(\kappa^{-1})\ .
\end{equation}
\\
The asymptotic behavior of the solutions $\Psi_{\pm}$ and $\Psi_{3}$  then turns out to be

\begin{equation}\label{4.27.1}
\Psi_{\pm}\skappa  
\frac{2\,\sqrt{\pi \lambda}}{K_{0}(\mu)}\ \Psi_{\WH}^{0}\,\frac{1}{\kappa_{1, 2}}\,
\left \{1-2\,\frac{\kappa_{2, 1}\, \kappa_{3}}{\kappa_{1, 2}^{3}} 
\right \}\,\exp \left \lbrack \frac{1}{\lambda}\, \left (
\kappa_{1, 2}^{2}-2\, \frac{\kappa_{2, 1}\, \kappa_{3}}{\kappa_{1, 2}}
\right )\,\right \rbrack 
\end{equation}
\\
and
\begin{equation}\label{4.27.2}
\Psi_{3}\skappa  
\frac{2\,\sqrt{\pi \lambda}}{K_{0}(\mu)}\ \Psi_{\WH}^{0}\,\frac{1}{\kappa_{3}}\,
\left \{1-2\,\frac{\kappa_{1} \kappa_{2}}{\kappa_{3}^{3}} 
\right \}\,\exp \left \lbrack \frac{1}{\lambda}\, \left (
\kappa_{3}^{2}-2\, \frac{\kappa_{1} \kappa_{2}}{\kappa_{3}}
\right )\,\right \rbrack\ . 
\end{equation}
\\
Since $f(z,\vec \kappa)$ is invariant under the permutation $\kappa_{1} \leftrightarrow \kappa_{2}$, the corresponding results for the case $\kappa_{1}<\kappa_{2}$ can now easily be obtained by exchanging the $\kappa_{j}$-indices $1 \leftrightarrow 2$ in the equations (\ref{4.27})-(\ref{4.27.2}). The asymmetric states $\Psi_{1}$ and $\Psi_{2}$ defined in (\ref{3.27}) then have to be constructed from $\Psi_{\pm}$ by using (\ref{3.28}). Consequently, the asymptotic behavior  of $\Psi_{i}, i\,\epsilon\, \{1,2,3 \}$, may be written in the closed form  

\begin{equation}\label{4.28}
\Psi_{i}\skappa  
\frac{2\,\sqrt{\pi \lambda}}{K_{0}(\mu)}\ \Psi_{\WH}^{0}\,\frac{1}{\kappa_{i}}\,
\left \{1-2\,\frac{\kappa_{j} \kappa_{k}}{\kappa_{i}^{3}} 
\right \}\,\exp \left \lbrack \frac{1}{\lambda}\, \left (
\kappa_{i}^{2}-2\, \frac{\kappa_{j} \kappa_{k}}{\kappa_{i}}
\right )\,\right \rbrack\ , \ \ 
\mbox{where}\ \ \varepsilon_{i j k}=1\ .
\end{equation}
\\
We should mention that these solutions diverge badly in the limit $\kappa \to \infty$, namely like $e^{\Lambda  a^{4}}$, so they are surely not normalizable in minisuperspace for any sensible choice of the scalar product. That is why we reject them as candidates for the physical quantum state of the Universe, and we are left with a just two dimensional, physical space of solutions spanned by $\Psi_{\Vil}$ and $\Psi_{\Vil}^{*}$.

\subsection{The no-boundary state}

In this section we will show that there exists a superposition of $\Psi_{\Vil}$ and $\Psi_{\Vil}^{*}$, such that this wavefunction is normalizable in minisuperspace {\em and} square-integrable on the caustic. Moreover, this uniquely determined solution will turn out to satisfy the no-boundary condition proposed by {\em Hartle} and {\em Hawking}  \cite{15,16}  (at least in the sense that the classical Universes described by this wavefunction are regular at $a=0$)
\footnote{A qualitative discussion of a no-boundary state for the anisotropic Bianchi IX metrics was first given in \cite{19} and, more explicitely, including a numerical plot for the wavefunction in the semi-classical limit, in \cite{20}. An expansion for small anisotropy has already been given in \cite{28}. However, the results obtained there are very lengthy and hard to interpret. For a discussion of the semi-classical trajectories generated by the no-boundary state in Ashtekar's variables see also \cite{24}}.
One may construct this solution by normalizing the Vilenkin state to approach unity in the limit $\beta_{+} \to + \infty$ at $\beta_{-}=0$ on the caustic. If one then considers the difference of this new  Vilenkin solution and its conjugate complex solution, the obtained distribution on the caustic is obviously square-integrable with respect to $\bpm$ (for further   explanation of this construction cf. fig.$\!$ \ref{grkv}). Finally, we shall choose the still unspecified overall normalization factor as usual in accordance to our convention (\ref{3.25}). The solution defined by this procedure turns out to be

\begin{equation}\label{4.29}
\Psi_{\NB}:=\Psi_{\WH}^{0}\ \Im \left({\cal N}_{\NB}\,
\int\limits_{\CC_{\Vil}} \dd u \ \exp\left\lbrack\, \frac{1}{\lambda}\,f(\sin u,\vec \kappa)\,\right\rbrack\,\right )\ ,
\end{equation}
\\
where
\begin{equation}\label{4.29+}
{\cal N}_{\NB}=2\,e^{\mu}\,\frac{
\II_{0}(\nu)-i\, \KK_{0}(\nu)}
{K_{0}(\mu)\, \II_{0}(\nu)-2 \pi\, I_{0}(\mu)\, \KK_{0}(\nu)}\qquad ,\ \ 
\mu=\frac{1}{2\,\lambda}\ ,\ \nu:=\frac{4}{\lambda}\ .
\end{equation}
\\
Here the special integrals

\begin{equation}\label{4.29++}
\II_{0}(\nu):=
\int\limits_{0}^{\frac{\pi}{2}}\,\dd x \,\exp \left \lbrack 
-\nu \sin^{4}x \right \rbrack
\ \ ,\ \
\KK_{0}(\nu):=
\int\limits_{0}^{\infty}\,\dd x \,\exp \left \lbrack
-\nu \cosh^{4}x \right \rbrack 
\end{equation}
\\
have been introduced which, to a certain extent, may be considered as generalized modified Bessel functions.\footnote{Similar normalization integrals occur in the calculation of ${\cal N}_{\Vil}$ and ${\cal N}_{0}$ in (\ref{3.26}), but with {\em squared} trigonometric functions in the exponent. Such integrals lead to the modified Bessel functions $I_{0}$ and $K_{0}$, that alternatively may be expressed as hypergeometric functions of the $_{1\!}F_{1}$-type. As a generalization, the integrals (\ref{4.29++}) may be written in terms of generalized hypergeometric functions of the $_{2\!}F_{2}$-type, but the integral $\KK_{0}$ requires  logarithmic contributions (as $K_{0}$) which, as far as we know, have no special name in the $_{2\!}F_{2}$-case. That is why we prefer to deal with the integral representations themselves.}

It is clear from its construction that $\Psi_{\NB}$ is integrable on the caustic, normalizable in minisuperspace in the distributional sense (as $\Psi_{\Vil}$) and we shall show below that it satisfies the no-boundary condition for $\hbar \to 0$. Furthermore, $\Psi_{\NB}$ obviously is a real-valued wavefunction. The behavior of the no-boundary state for $\kappa \to \infty$ can be immediately extracted from the asymptotics of the Vilenkin state (\ref{4.21}). In this way  an asymptotic description of $\Psi_{\NB}$ in the Lorentzian regime is available, so we will restrict ourselves to the Euclidean regime throughout the following. To discuss the limits $\hbar \to 0$ and $\lambda \to 0$ we shall first expand the normalization factor ${\cal N}_{\NB}$ for small $\lambda$.   Saddle-point expansions of $\II_{0}$ and $\KK_{0}$ in the corresponding limit $\nu \to \infty$ finally yield for the leading term in the asymptotic series

\begin{equation}\label{4.29.1}
{\cal N}_{\NB} \slambda \frac{2\,e^{\frac{1}
{\lambda}}}{\sqrt{\pi \lambda}} \ \slambda i\, {\cal N}_{0}\ ,
\end{equation}
\\
where we have also used (\ref{3.26}). Obviously  the normalization factors ${\cal N}_{\NB}$ and ${\cal N}_{0}$ have the same asymptotic behavior in the limit $\lambda \to 0$, where $\Psi_{\NB}$ may be written in the form

\begin{equation}\label{4.29.2}
\Psi_{\NB} \slambda \frac{1}{2}\ \Psi_{\WH}^{0}\ \, {\cal N}_{0}\!\!\!\! 
\!\!\int \limits_{\CC_{\Vil} \ominus \CC_{\Vil}^{*}}\,
\!\!\!\!\!\!\dd u \ \exp\left\lbrack\, \frac{1}{\lambda}\,f(\sin u,\vec \kappa)\,\right\rbrack\ .
\end{equation}
\\
If one now chooses the integration path  $\CC_{-} \oplus \CC_{+} \oplus \CC_{3} \ominus \CC_{0}$, which is equivalent to $\CC_{\Vil} \ominus \CC_{\Vil}^{*}$, the final expression for the asymptotic behavior in the limit $\lambda \to 0$ becomes: 

\begin{equation}\label{4.30}
\Psi_{\NB} \slambda \frac{1}{2} \left( \, \sum_{j=1}^{3}\, \Psi_{j} \,- \Psi_{0} \,\right )\ .
\end{equation}
\\
This representation displays nicely the individual saddle-point contributions in the Euclidean regime and will prove useful for the following discussions. 

\subsubsection{Semi-classical limit $\hbar \to 0$}

\noindent
Since for $\hbar \to 0$ the saddle-point $z_{0}$ always provides the dominating contribution in comparison with $z_{\pm}$ and $z_{3}$, the relation (\ref{4.30}) implies 

\begin{equation}\label{4.31}
\Psi_{\NB} \shbar -\frac{1}{2}\,\Psi_{0},
\end{equation}
\\
and all that remains to be considered is an expansion of $\Psi_{0}$ in the limit $\hbar \to 0$.

Using once more the expansion (\ref{4.2}), the action of the solution $\Psi_{0}$ is calculated to be

\begin{equation}\label{4.32}
S_{0}^{(0)}=- \frac{3 \pi i}{\Lambda}\, \left \lbrack
1- \left (1 - \left (\frac{a}{a_{0}} \right )^{2} \right )^{\frac{3}{2}} \right \rbrack
\end{equation}
\\
in the isotropic case, where also the normalization factor ${\cal N}_{0}$ has been taken into account (in contrast to defintion (\ref{4.3})). The expression is easily recognized as the negative Vilenkin action (\ref{4.5}). Consequently the classical trajectories and spacetime metrics are the same as in the Vilenkin case up to a reversal of the $\tau$-direction and, as there, the no-boundary condition is satisfied in the isotropic case. 
 
But let us now consider the influence of anisotropy: Then the action is of the form

\begin{equation}\label{4.33}
S_{0}=S_{0}^{(0)}+36 \pi i \,a^2\,
\frac{3-\sqrt{\,1-\frac{1}{3}\,a^{2} \Lambda } }
{24+a^{2} \Lambda }
\, (\beta_{+}^{\, 2}+\beta_{-}^{\,2})+{\cal O}(\beta_{\pm}^{\, 3})\ ,
\end{equation}
\\
implying (with $\kappa=\frac{1}{12}\,\Lambda a^{2}$)

\begin{equation}\label{4.34}
\frac{\dd a}{\dd \tau} \approx 
\sqrt{\ 1-4\kappa}-4\ \frac{\beta_{+}^{\,2}+\beta_{-}^{\,2}}{(\kappa+2)^{2}}\, 
\left \lbrack\, 
3+\frac{(\kappa+3)^2-10}{\sqrt{\ 1- 4 \kappa}} \,\right \rbrack
\  ,
\end{equation}

\vspace{0.2 cm}

\begin{equation}\label{4.35}
\frac{\dd \beta_{\pm}}{\dd \tau}
\approx \frac{2}{a}\  \frac{3-\sqrt{\ 1- 4 \kappa}}{2+\kappa}\,
\beta_{\pm}
\end{equation}
\\
for the Euclidean, classical trajectories. As the prefactor of $\bpm$ in  (\ref{4.35}) is positive definite, the point $\bpm=0$ is attractive for decreasing $\tau$, i.e. in this $\tau$-direction $\beta$ tends to zero. Then in (\ref{4.34}) the $\bpm$-term may be neglected, and the scale factor will reach $a=0$ at a {\em finite} value of $\tau$, say $\tau=0$. The asymptotic form of (\ref{4.34}) and (\ref{4.35}) for $\tau \to 0$ simply reads

\begin{equation}\label{4.36}
\frac{\dd a}{\dd \tau} \ \stau 1\qquad ,\qquad 
\frac{\dd \beta_{\pm}}{\dd \tau}\stau \frac{2}{a}\  
\beta_{\pm}\ ,
\end{equation}
\\
and the $\bpm$-equation can be integrated to give

\begin{equation}\label{4.37}
\frac{\dd}{\dd \tau}\, \mbox{ln} \, \bpm \stau \frac{2}{a} \, 
\frac{\dd a}{\dd \tau}=\frac{\dd}{\dd \tau}\, \mbox{ln} \,a^{2}
\qquad \Leftrightarrow \qquad \bpm \ptau a^{2}\ .
\end{equation}
\\
Consequently, the classical Universes become exactly isotropic in reaching $a=0$, and there look the same as in the case $\bpm \equiv 0$. Therefore also Universes anisotropic at $a > 0$ remain   regular as $a$ tends to zero. So $\Psi_{\NB}$ is indeed a solution of the Wheeler-DeWitt equation which satisfies the no-boundary proposal semi-classically and its name is justified, after all.

If (\ref{4.34}) and (\ref{4.35}) are considered with increasing $\tau$, the anisotropy grows exponentially as in the Vilenkin case, and the validity of these equations breaks down. Finally, the trajectories approach the caustic, and tunneling processes to Lorentzian space-time become more and more probable.

\subsubsection{The limit $\Lambda \to 0$}

To discuss the limit $\Lambda \to 0$ of the no-boundary state $\Psi_{\NB}$ we will again make use of the relation (\ref{4.30}). Since $\Psi_{i} \to \Psi_{i}^{0}$ for $\Lambda \to 0$ is known from before for $i\,\epsilon\, \{1, 2, 3 \}$, only the behavior of $\Psi_{0}$ in (\ref{4.30}) remains to be discussed. An expansion of $\Psi_{0}$ for small $\Lambda$ is straightforward and we just give the final results:

\begin{equation}\label{4.38}
z_{0}=1-\frac{\lambda}{4}\,(\sigma_{1}+\sigma_{2})+{\cal O}
(\lambda^{2})\ ,
\end{equation}

\begin{equation}\label{4.39}
S_{0}=i\,\biggl \lbrack\,\frac{6 \pi}{\Lambda}+ \Phi - \hbar\,(\sigma_{1} + \sigma_{2}+\sigma_{3}) \,\biggr \rbrack+{\cal O}(\Lambda)\ ,
\end{equation}

\begin{equation}\label{4.40}
A_{0}=\frac{i}{2}\,\biggl \lbrack\, 1 +\frac{ \lambda}{4}\,(\sigma_{1}+\sigma_{2}+\sigma_{3}) \,\biggr \rbrack+{\cal O}(\Lambda^{2})\ ,
\end{equation}

\begin{equation}\label{4.41}
\lim_{\Lambda \to 0}\, \Psi_{0} =
\Psi_{\WH}^{0}\,\exp \Bigl \lbrack
\sigma_{1}+\sigma_{2}+\sigma_{3} \Bigr \rbrack =\Psi_{\NB}^{0}\ .
\end{equation}
\\
Obviously the no-boundary state (\ref{1.25}) of the $\Lambda=0$ -model is approached by $\Psi_{0}$. From equation (\ref{4.30}) it now follows, using (\ref{4.26}) and (\ref{4.41})

\begin{equation}\label{4.41.1}
\lim_{\Lambda \to 0}\, \Psi_{\NB} =\frac{1}{2} \, \left (\, \sum_{j=1}^{3} \Psi_{j}^{0}-\Psi_{\NB}^{0} \right )\ .
\end{equation}
\\
We can see that $\Psi_{\NB}$ does {\em not} approach the no-boundary state $\Psi_{\NB}^{0}$ of the $\Lambda =0$ -model when $\Lambda$ tends to zero even though both wavefunctions (as many others!) satisfy the no-boundary condition semi-classically for $\hbar \to 0$. The difference comes from our additional requirements that $\Psi_{\NB}$ for $\Lambda \not= 0$ be normalizable  in the distributional sense with respect to all three variables $\{\alpha, \bpm\}$ {\em and} be square-integrable with respect to $\bpm$ on the caustic\footnote{The property of normalizability in minisuperspace opens again the possibility to define a scalar product for physical states as in \cite{27}, and to consider the no-boundary state as a ground state in the resulting Hilbert space.}. The normalizability of $\Psi_{\NB}$ for $a \to \infty$, like that of $\Psi_{\Vil}$, is directly related to the fact that, semi-classically, these states describe a Lorentzian Universe in that limit. E.g. had we defined the no-boundary state simply by $\Psi_{0}$ this would have given an acceptable limit $\Psi_{\NB}^{0}$ for $\Lambda \to 0$, but for $\Lambda \not= 0$ the semi-classical limit of $\Psi_{0}$ gives, besides a Lorentzian, also Euclidean contributions.
As a consequence, this alternative ``no-boundary state'' would contain components, which describe additional Euclidean Universes for $a \to \infty$, and make the wavefunction diverge in this limit. Furthermore, it is easily checked that $\Psi_{0}$ is a wavefunction which is {\em not} square-integrable on the caustic.  

\begin{figure}
\begin{center}
\hskip 0 cm
\psfig{figure=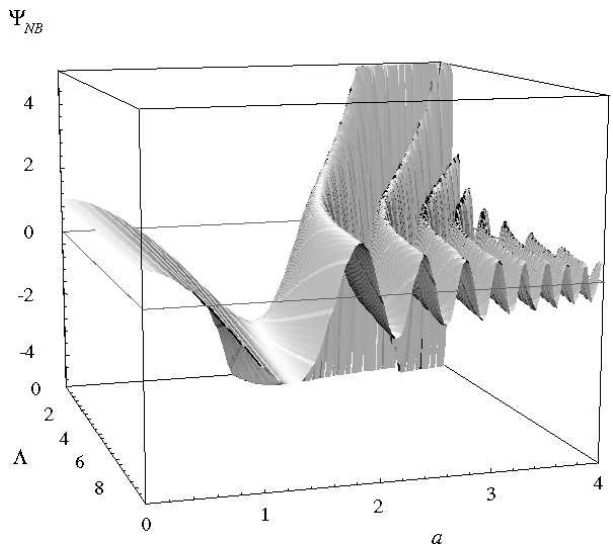,width=8cm}
\end{center}
\caption{The no-boundary state}\label{granb}
\end{figure}

\noindent
Figure \ref{granb} shows a numerical plot of the no-boundary state in the $\{ a, \Lambda \}$-plane analogous to fig.$\!$ \ref{gravil}. To allow a clear view the plot has been bounded to values of $\Psi_{\NB}$ lying in the interval $\lbrack -5, +5 \rbrack$. Like in the Vilenkin case, rapid oscillations start in the Lorentzian regime, and clearly $\Psi_{\NB} \to 1$ as $a \to 0$. In view of the strong increase of the amplitude with $a$ at small $\Lambda$, one might worry about normalizability in minisuperspace.  However, in connection with (\ref{4.29}) the asymptotic result for the Vilenkin state (\ref{4.21}) guarantees that $\Psi_{\NB}$ falls off as $a^{- \frac {3}{2}}$ for {\em each} $\Lambda$ if sufficiently large scale factors are considered.

\begin{figure}
\begin{center}
\hskip 0 cm
\psfig{figure=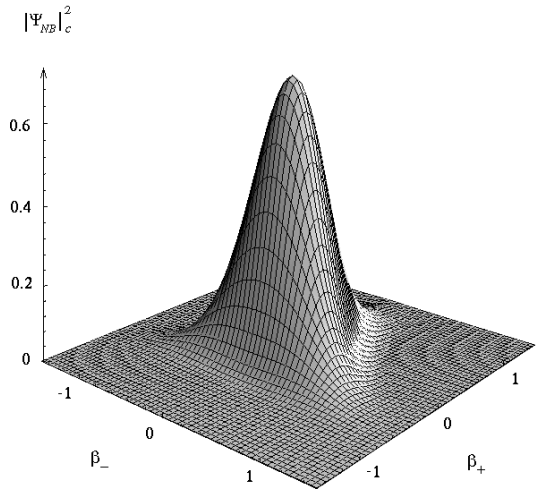,width=7cm}
\end{center}
\caption{The initial value distribution generated by the no-boundary state}\label{grknb}
\end{figure}

\noindent
Finally, fig.$\!$ \ref{grknb} shows the resulting initial value distribution $|\Psi_{\NB}|^{2}_{c}$ for the classical evolution of the Universe. As announced above, $|\Psi_{\NB}|^{2}_{c}$ rapidly approaches zero with increasing $\beta$ in all directions of the $\bpm$-plane, thus representing an integrable probability distribution. The figure shows that the quantum state $\Psi_{\NB}$ implies an {\em isotropic} classical Universe in the Lorentzian regime, apart from possible quantum fluctuations about $\bpm=0$.

\section{Conclusion}\label{5}

The central purpose of this paper was to derive exact solutions of the Wheeler-DeWitt equation with cosmological constant by employing a generalized Fourier-transformation to the Chern-Simons solution (\ref{2.6}). As a result we found five linearly independent states, which approach the five known $\Lambda=0$ -states when $\Lambda$ tends to zero. However, for $\Lambda \not= 0$, there is just a two dimensional subspace which appears to be physically relevant in the sense that the states are normalizable wavefunctions in minisuperspace and allow a classical Universe with pseudo-Riemannian geometry. We were able to construct a basis of this subspace in form of the solutions $\Psi_{\Vil}$ and $\Psi_{\NB}$, which, semi-classically, satisfy the boundary conditions proposed by {\em Vilenkin} and {\em Hartle and Hawking}, respectively. Moreover, with the additional requirement of integrability on the caustic, just one solution remains, namely the no-boundary state $\Psi_{\NB}$. We may add a few remarks comparing the two physically interesting solutions $\Psi_{\NB}$ and $\Psi_{\Vil}$.

If the Euclidean geometries are considered, which arise from the semi-classical trajectories in the Euclidean domain of minisuperspace by introducing an imaginary time variable $\dd \tau = i\,\dd t$, the two solutions behave very differently: While a Universe described by the no-boundary state becomes isotropic and remains regular in approaching $a=0$, the Vilenkin-trajectories diverge in some $\bpm$-directions for $a \to 0$, i.e. the corresponding Euclidean 4-geometry behaves singular for small scale parameters.

If one considers the Euclidean approach to the caustic then all trajectories generated by the Vilenkin state reach the caustic in the isotropic point $\kappa=\frac{1}{4}, \bpm=0$, whereas the no-boundary trajectories run to infinite anisotropy.

The creation of a classical Universe with pseudo-Riemannian geometry happens by a tunneling process through this caustic. The tunneling probability is given as the value of $|\Psi|^{2}_{c}$ on the caustic and thus is a function of $\bpm$ only ($\Lambda$ taken as fixed). The results for these distributions, which rule the initial values for the classical evolution of the Universe,  are shown in fig.$\!$ \ref{grkv} and fig.$\!$ \ref{grknb} and predict both an isotropic Universe as the most probable classical Universe among all alternatives. However, just the no-boundary distribution is sharply concentrated about $\bpm=0$ and integrable with respect to the anisotropy variables, whereas the Vilenkin distribution is broader and {\em not} integrable (even though bounded).

In the Lorentzian regime both wavefunctions describe a DeSitter-like Universe, which becomes more and more isotropic as $a$ tends to infinity. As a difference, $\Psi_{\NB}$ always describes expanding and collapsing Universes simultaneously, while $\Psi_{\Vil}$ just describes the expanding phase, and thus is not invariant under time reversal.

In summary one might prefer the no-boundary solution $\Psi_{\NB}$ as {\em  the} quantum state of the Universe. However, we should stress the fact that it is not the no-boundary idea itself which picks out uniquely this solution among the five solutions described by the Chern-Simons functional, but several additional, physically well-motivated integrability conditions have to be imposed on the solution to get $\Psi_{\NB}$. Then, the resulting wavefunction {\em is found}  to satisfy the no-boundary proposal {\em in the semi-classical limit} $\hbar \to 0$.

\vspace{0.3 cm}

Finally, it may be of interest to consider the application of the results obtained here to a model with a massive, scalar matter-field\footnote{For solutions to the isotropic model see \cite{2,3,14,15,16}.}. We want to show in brief that the $\Lambda$-solutions found above are directly applicable to this model in the limit $m \to 0$:

A massive Klein-Gordon field $\phi$ gives a contribution

\begin{equation}\label{5.1}
{\cal S}_{mat}\, \lbrack g_{\mu \nu}, \phi \rbrack=\ -\frac{1}{8 \pi}\ \int\limits_{{\cal M}} \dd^{4} x \sqrt{-g}\ \Bigl (g^{\mu \nu} \phi_{,\mu}\phi_{,\nu}+m^{2} \phi^{2} \Bigr )
\end{equation}
\\
to the action (\ref{1.1}) and finally yields an additive term 

\begin{equation}\label{5.2}
H_{mat}=3\, \Bigl \lbrack - \hbar^{2}\, \partial_{\phi}^{2}+\pi^{2} a^{6} m^{2} \phi^{2} \Bigr \rbrack
\end{equation}
\\
in the Hamiltonian (\ref{1.15}). If now the field amplitude $\phi$ is rescaled with the mass $m$, $\eta :=m\, \phi$, $H_{mat}$ is converted to

\begin{equation}\label{5.3}
H_{mat}=3\,\Bigl \lbrack -\hbar^{2} m^{2} \partial_{\eta}^{2}+\pi^{2} a^{6} \eta^{2} \Bigr \rbrack \ .
\end{equation}
\\
Comparing with (\ref{1.20}) it is easily seen that in the limit $m \to 0$ this term corresponds exactly to the contribution which arises from a cosmological constant $\Lambda = \eta^{2}$. Thus one obtains the asymtotic solution for the Bianchi type IX model with a massive, scalar field

\begin{equation}\label{5.4}
\Psi(\alpha,\bpm,\phi;m)\, \sm \Psi(\alpha,\bpm;\Lambda=m^{2} \phi^{2})
\end{equation}
\\ 
at fixed $\Lambda = m^{2} \phi^{2}$, which, by further expansion about $m=0$, may be extended to small, but non-vanishing masses of the field.

\vspace{0.3 cm}

As an interesting project for future consideration we leave the application of the generalized Fourier-transformation to the general form of the Chern-Simons functional \cite{9,10}. It would be interesting to determine whether again a Vilenkin and a no-boundary state are obtained as topologically inequivalent functional integrals from the Chern-Simons solutions. Work in this direction is in progress.

\acknowledgements

Support of this work by the Deutsche Forschungsgemeinschaft through the Sonderforschungsbereich ``Unordnung und gro{\ss}e Fluktuationen'' is gratefully acknowledged.


\begin{thebibliography}{99}
  
\bibitem {1} R.~Arnowitt, S.~Deser and C.~W.~Misner, in {\em Gravitation, An Introduction to Current Research}, ed. L.~Witten (John Wiley and Sons 1962), Chap.7 

\bibitem {2} J.~A.~Wheeler, in {\em Relativity and Topology}, ed. C.~DeWitt and B.~S.~DeWitt (Gordon and Breach, New York 1964)

\bibitem {3} B.~S.~DeWitt, Phys.~Rev.~D {\bf 16}, 1113 (1967)

\bibitem {4} A.~Ashtekar, Phys.~Rev.~Lett. {\bf 57}, 2244 (1986);
Phys.~Rev.~D {\bf 36}, 295 (1987)

\bibitem {5} A.~Ashtekar, in {\em Lectures on Nonperturbative Canonical Gravity} (World Scientific, Singapore 1991)

\bibitem {6} A.~Ashtekar, in {\em Polymer Geometry at the Planck Scale and the Quantum Einstein Equations}, electronic archive {\em hep-th/9601054}

\bibitem {7} C.~J.~Isham, {\em Structural Issues in Quantum Gravity}, electronic archive {\em gr-qc/9510063}

\bibitem {8} K.~Ezawa, {\em Nonperturbative solutions for canonical quantum gravity: an overview}, electronic archive {\em gr-qc/9601050}

\bibitem {9} H.~Kodama, Phys.~Rev.~D {\bf 42}, 2548 (1990)

\bibitem {10} M.~P.~Brencowe, Nucl.~Phys.~B {\bf 341}, 213 (1990)

\bibitem {11} R.~Graham, Phys.~Rev.~Lett. {\bf 67}, 1381 (1991)

\bibitem {12} J.~Bene and R.~Graham, Phys.~Rev.~D {\bf 49},799 (1994)

\bibitem {12a} R.~Graham and H.~Luckock, Phys.~Rev.D {\bf 49}, R4981 (1994)

\bibitem {13} A.~Csord\'as and R.~Graham, Phys.~Rev.~Lett. {\bf 74}, 4129 (1995)

\bibitem {14} A.~Vilenkin, Phys.~Rev.~D {\bf 33}, 3560 (1986)
 
\bibitem {15} J.~B.~Hartle und S.~W.~Hawking, Phys.~Rev.~D {\bf 28}, 2960 (1983)

\bibitem {16} S.~W.~Hawking, Nucl.~Phys. B {\bf 239}, 257 (1984)

\bibitem {17} C.~W.~Misner, K.~S.~Thorne und J.~A.~Wheeler, in  {\it Gravitation.} (W.~H.~Freeman and Company, San Francisco 1973)

\bibitem {18} M.~P.~Ryan, Jr.~, und L.~C.~Shepley, in {\it Homogeneous Relativistic Cosmologies.} (Princeton University Press, Princeton 1975)

\bibitem {19} S.~W.~Hawking und J.~C.~Luttrell, Phys.~Lett.~B {\bf 143}, 
83 (1984)

\bibitem {20} W.~A.~Wright I.~G.~Moss, Pys.~Lett.~B {\bf 154}, 115

\bibitem {21} S.~Del Campo und A.~Vilenkin, Phys.~Lett.~B {\bf 224}, 45 
(1989)

\bibitem {22a} R.~Graham and P.~Sz\'epfalusy, Phys.~Rev.~D {\bf 42}, 2483 (1990)

\bibitem {22} V.~Moncrief and M.~P.~Ryan, Jr., Phys.~Rev.~D {\bf 44}, 2375 (1991)

\bibitem {23} A.~Ashtekar and G.~Pullin, Ann.~Isr.~Phys.~Soc. {\bf 9}, 65 (1990)

\bibitem {24} J.~Louko, Phys.~Rev.~D {\bf 51}, 586 (1995)
  
\bibitem {25} R.~Courant und D.~Hilbert, in {\it Methoden der mathematischen Physik I.} (Springer, Berlin 1968)

\bibitem {26} C.~Bender und S.~Orszag, in {\it Advanced mathematical methods for scientists and engineers.} (McGraw-Hill Book Company, Inc., New York 1978)

\bibitem {27} D.~Marolf, Class.~Quant.~Grav. {\bf 12}, 1199 (1995)

\bibitem {28} P.~Amsterdamski, Phys.~Rev.~D {\bf 31}, 3073 (1985)


\end{thebibliography}
\end{document}